\documentclass[aps,prl,reprint,twocolumn,superscriptaddress,floatfix,nofootinbib,longbibliography]{revtex4-1}
\usepackage{graphicx,amsmath,amsfonts,amssymb,amsthm,xr}
\usepackage{epsfig,amsmath,amssymb,color,dsfont,upgreek,physics}
\usepackage{mathrsfs}
\usepackage{mathtools}
\usepackage{bbold}
\usepackage{comment}
\usepackage{float}

\usepackage[bookmarks=true,colorlinks,linkcolor=OrangeRed,urlcolor=NavyBlue,citecolor=RoyalBlue]{hyperref}
\usepackage[dvipsnames]{xcolor}

\definecolor{mygold}{rgb}{0.93,0.69,0.13}

\definecolor{mypurple}{rgb}{0.49,0.18,0.56}

\newcommand{\mbeq}{\overset{!}{=}}\definecolor{mygreen}{rgb}{0,0.5,0}

\definecolor{mygreen}{rgb}{0,0.5,0}
\definecolor{myred}{rgb}{0.7,0,0}
\definecolor{myblue}{rgb}{0,0,0.5}

\begin{document}
\title{Temperature-Induced Disorder-Free Localization}
\author{Jad C.~Halimeh}
\email{jad.halimeh@physik.lmu.de}
\affiliation{Department of Physics and Arnold Sommerfeld Center for Theoretical Physics (ASC), Ludwig-Maximilians-Universit\"at M\"unchen, Theresienstra\ss e 37, D-80333 M\"unchen, Germany}
\affiliation{Munich Center for Quantum Science and Technology (MCQST), Schellingstra\ss e 4, D-80799 M\"unchen, Germany}
\author{Philipp Hauke}
\affiliation{INO-CNR BEC Center and Department of Physics, University of Trento, Via Sommarive 14, I-38123 Trento, Italy}
\affiliation{INFN-TIFPA, Trento Institute for Fundamental Physics and Applications, Trento, Italy}
\author{Johannes Knolle}
\affiliation{Department of Physics, Technische Universit\"at M\"unchen, James-Franck-Straße 1, D-85748 Garching, Germany}
\affiliation{Munich Center for Quantum Science and Technology (MCQST), Schellingstra\ss e 4, D-80799 M\"unchen, Germany}
\affiliation{Blackett Laboratory, Imperial College London, London SW7 2AZ, United Kingdom}
\author{Fabian Grusdt}
\affiliation{Department of Physics and Arnold Sommerfeld Center for Theoretical Physics (ASC), Ludwig-Maximilians-Universit\"at M\"unchen, Theresienstra\ss e 37, D-80333 M\"unchen, Germany}
\affiliation{Munich Center for Quantum Science and Technology (MCQST), Schellingstra\ss e 4, D-80799 M\"unchen, Germany}
\begin{abstract}
Disorder-free localization is a paradigm of strong ergodicity breaking that has been shown to occur in global quenches of lattice gauge theories when the system is initialized in a superposition over an extensive number of gauge sectors. Here, we show that preparing the system in a thermal Gibbs ensemble without any coherences between different gauge sectors also gives rise to disorder-free localization, with temperature acting as a disorder strength. We demonstrate our findings by calculating the quench dynamics of the imbalance of thermal ensembles in both $\mathrm{U}(1)$ and $\mathbb{Z}_2$ lattice gauge theories through exact diagonalization, showing greater localization with increasing ensemble temperature. Furthermore, we show how adding terms linear in local pseudogenerators can enhance temperature-induced disorder-free localization due to the dynamical emergence of an enriched local symmetry. Our work expands the realm of disorder-free localization into finite-temperature physics, and shows counterintuitively that certain quantum nonergodic phenomena can become more prominent at high temperature. We discuss the accessibility of our conclusions in current quantum simulation and computing platforms.
\end{abstract}

\date{\today}
\maketitle

Recent years have witnessed the introduction of an array of exotic quantum many-body phenomena showing nonergodic behavior such as many-body localization \cite{Basko2006,Alet_review,Abanin_review}, quantum scars \cite{Moudgalya2018,Turner2018}, and Hilbert-space fragmentation \cite{Sala2020,Khemani2020}. They not only challenge the Eigenstate Thermalization Hypothesis (ETH) \cite{Deutsch_review,Rigol_review} by delaying thermalization up to significant times or avoiding it altogether, but they also promise to be of practical importance in quantum information technology \cite{Huse2013,Bauer2013,Yao2015}, and can give rise to fascinating phases of quantum matter with no equilibrium counterpart such as discrete time crystals \cite{vonKeyserlingk2016,Khemani2016,Else2016,Yao2017,Mi2022}. A particularly intriguing phenomenon in this vein is that of disorder-free localization (DFL) in lattice gauge theories (LGTs) \cite{Smith2017,Brenes2018,smith2017absence,Metavitsiadis2017,Smith2018,Russomanno2020,Papaefstathiou2020,karpov2021disorder,hart2021logarithmic,Zhu2021,Sous2021}. Despite the absence of quenched disorder in the system, DFL can still arise in LGTs in the wake of quenching simple product states that comprise a superposition over an extensive number of gauge superselection sectors \cite{Smith2017,Brenes2018}, which act as an effective disorder background leading to localization \cite{Smith2018}.

\begin{figure}[t!]
    \centering
    \includegraphics[width=\columnwidth]{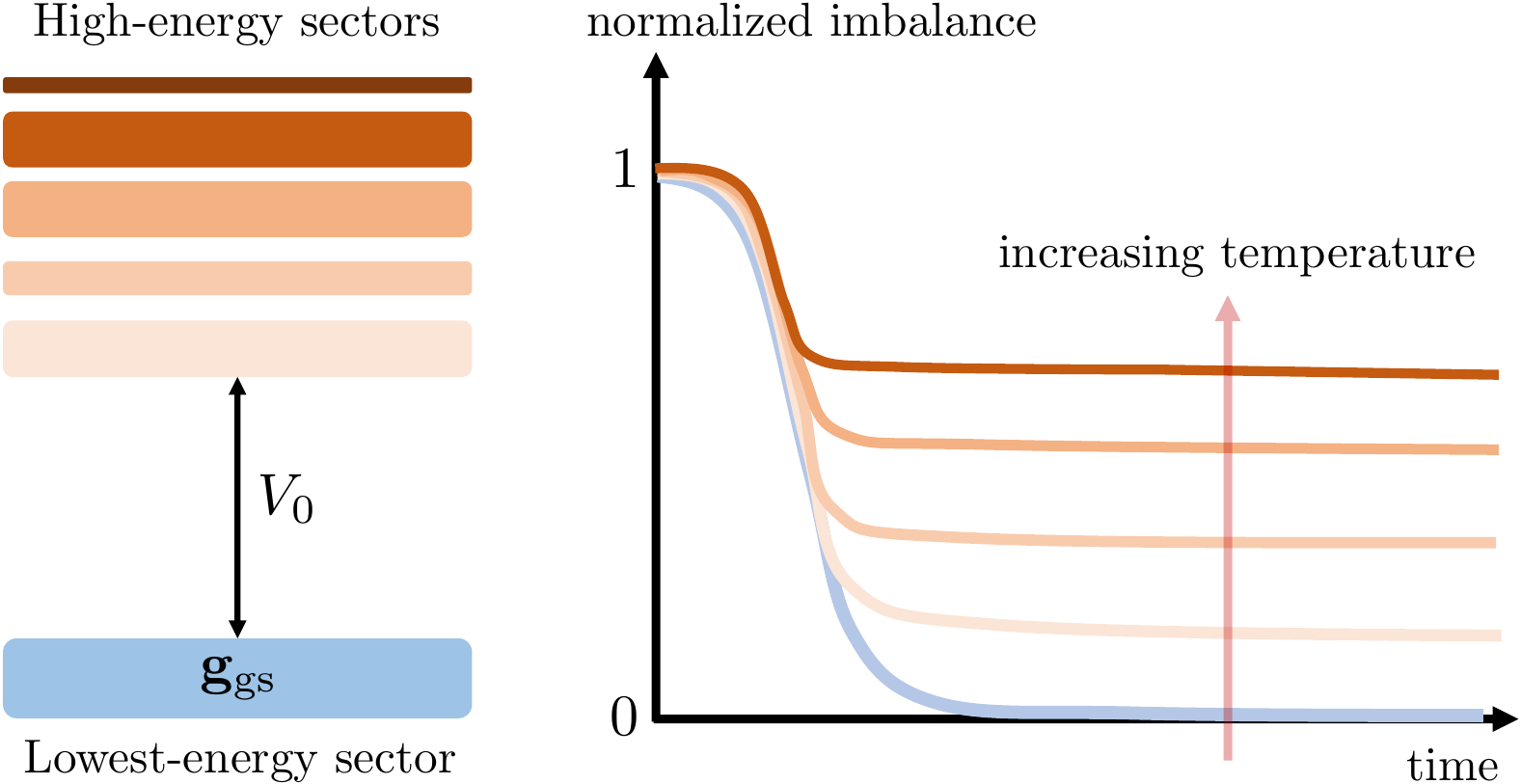}
    \caption{(Color online). A preparation Hamiltonian is used to prepare a system in a thermal ensemble, which is subsequently quenched to a LGT Hamiltonian. At zero temperature, the ensemble is a pure domain-wall state in a homogeneous gauge sector $\mathbf{g}_\text{gs}$, which is gapped by $V_0$ from other sectors, and the imbalance decays to zero at long times. At sufficiently high temperatures, the ensemble comprises nonnegligible weights over an extensive number of gauge sectors, which leads to the dynamical emergence of an effective disorder over their associated background charges, leading to localization in the form of a finite nonzero plateau in the imbalance at long times.}
    \label{fig:schematic}
\end{figure}

A major incentive for building quantum simulation and computing platforms is the promise to simulate quantum many-body models and explore their rich physics \cite{Cirac2012,Hauke2012,Alexeev_review,klco2021standard}. Although most experimental works have been focused on ground-state properties and dynamics of pure quantum states, simulating finite-temperature physics is crucial for a multitude of applications such as, e.g., high-$T_\mathrm{c}$ superconductivity, thermal phase transitions, and finite-temperature (de)confinement. Recently, there has been several experimentally relevant proposals for probing finite-temperature physics on quantum simulators and computers \cite{Wu2019,Sagastizabal2021,Zhang2021CV,Lu2022,Schuckert2022}. This motivates studying phenomena such as DFL at finite temperature, and probing whether localization can persist away from the paradigm of a pure initial state.

In this work, we show that quenching finite-temperature thermal ensembles in LGTs leads to nonergodic dynamics even when the initial ensemble has no coherences between the different gauge sectors; cf.~Fig.~\ref{fig:schematic}. Rather counterintuitively, this \textit{temperature-induced DFL} ($T$-DFL), which is ultimately a quantum interference effect, becomes more prominent with increasing ensemble temperature. We showcase our findings in two paradigmatic LGTs that have been the focus of many recent experimental works, and also highlight how the use of local pseudogenerators \cite{Halimeh2021stabilizing} enhances $T$-DFL due to the dynamical emergence of an enriched local symmetry \cite{Halimeh2021enhancing}.

\textbf{\textit{Models.---}}The principal property of a LGT is its gauge symmetry \cite{Rothe_book,Zee_book}, which is introduced by the local generators $\hat{G}_j$ and encoded in the commutation relations $\big[\hat{H}_0,\hat{G}_j\big]{=}0,\,\forall j$, where $j$ denotes a site in a system of size $L$. Matter fields live on these sites, while gauge and electric fields reside on the links in between. The local generator $\hat{G}_j$ imposes an intrinsic local relationship between the matter occupation on site $j$ and the configurations of the electric fields on its adjacent links. Gauge-invariant states $\ket{\phi}$ are simultaneous eigenstates of all generators: $\hat{G}_j\ket{\phi}{=}g_j\ket{\phi}$. The eigenvalues $g_j$ are so-called background charges, and a set for the entire system of $L$ sites defines a unique gauge superselection sector $\mathbf{g}{=}(g_1,g_2,\ldots,g_L)$. We will add an appropriate superscript to $\hat{H}_0$ and $\hat{G}_j$ when referring to a specific LGT, but leave them without such a superscript when the discussion is general. Periodic boundary conditions are employed throughout this work.

The $\mathrm{U}(1)$ quantum link model (QLM) is a formulation of lattice quantum electrodynamics where the infinite-dimensional gauge and electric fields are represented by spin-$S$ operators \cite{Chandrasekharan1997,Wiese_review,Hauke2013,Yang2016,Kasper2017}. Its Hamiltonian is given by
\begin{align}\nonumber
    \hat{H}_0^{\mathrm{U}(1)}=\sum_{j=1}^L\bigg[&\frac{J}{2\sqrt{S(S+1)}}\big(\hat{\sigma}^-_j\hat{s}^+_{j,j+1}\hat{\sigma}^-_{j+1}+\text{H.c.}\big)\\\label{eq:H0_U1}
    &+\frac{\mu}{2}\hat{\sigma}^z_j+\frac{\kappa^2}{2}\big(\hat{s}^z_{j,j+1}\big)^2\Big],
\end{align}
where the Pauli operator $\hat{\sigma}^z_j$ represents the matter field on site $j$, and the spin-$S$ operators $\hat{s}^{z}_{j,j+1}$ and $\hat{s}^+_{j,j+1}/\sqrt{S(S+1)}$ represent the electric and gauge fields, respectively, on the link between sites $j$ and $j{+}1$. The number of sites is denoted by $L$, the fermionic mass by $\mu$, the gauge coupling by $\kappa$, and $J{=}1$ sets the overall energy scale. The local generator of the $\mathrm{U}(1)$ gauge symmetry is given by
\begin{align}\label{eq:Gj_U1}
    \hat{G}_j^{\mathrm{U}(1)}=(-1)^j\big(\hat{n}_j+\hat{s}^z_{j-1,j}+\hat{s}^z_{j,j+1}\big),
\end{align}
where $\hat{n}_j{=}(\hat{\sigma}^z_j{+}\mathds{1})/2$. The gauge invariance of Eq.~\eqref{eq:H0_U1} is encapsulated in the commutation relations $\big[\hat{H}_0^{\mathrm{U}(1)},\hat{G}_j^{\mathrm{U}(1)}\big]{=}0,\,\forall j$.

The $\mathbb{Z}_2$ LGT is described by the Hamiltonian \cite{Zohar2017,Borla2019,Yang2020fragmentation,kebric2021confinement}
\begin{align}\label{eq:H0_Z2}
    \hat{H}_0^{\mathbb{Z}_2}=\sum_{j=1}^L\big[J\big(\hat{a}_j^\dagger\hat{\tau}^z_{j,j+1}\hat{a}_{j+1}+\text{H.c.}\big)-h\hat{\tau}^x_{j,j+1}\big],
\end{align}
where $\hat{a}_j,\hat{a}_j^\dagger$ are hard-core bosonic annihilation and creation operators with $\hat{n}_j{=}\hat{a}_j^\dagger\hat{a}_j$ representing matter occupation on site $j$, and the Pauli operator $\hat{\tau}^{x(z)}_{j,j+1}$ represents the electric (gauge) field on the link between sites $j$ and $j{+}1$. The generator of the $\mathbb{Z}_2$ symmetry is
\begin{align}
    \hat{G}^{\mathbb{Z}_2}_j=(-1)^{\hat{n}_j}\hat{\tau}^x_{j-1,j}\hat{\tau}^x_{j,j+1},
\end{align}
where the gauge invariance of Hamiltonian~\eqref{eq:H0_Z2} is manifest in the commutation relations $\big[\hat{H}^{\mathbb{Z}_2}_0,\hat{G}^{\mathbb{Z}_2}_j\big]{=}0,\,\forall j$.

Both models~\eqref{eq:H0_U1} and~\eqref{eq:H0_Z2} have recently been experimentally realized in synthetic quantum matter setups \cite{Martinez2016,Goerg2019,Schweizer2019,Yang2020,Zhou2021,Wang2022,Mildenberger2022}.

\textbf{\textit{Thermal ensemble and quench dynamics.---}}
Let us consider the \textit{preparation} Hamiltonian
\begin{align}\label{eq:prep}
\hat{H}_\mathrm{prep}=\sum_{j=1}^L\Big[V_0\big(\hat{G}_j-g^\mathrm{gs}_j\big)^2-\alpha_j\hat{n}_j\Big],
\end{align}
where $\alpha_j{=}+1$ for $j{\leq}L/2$ and $\alpha_j{=}-1$ for $j{>}L/2$, and $V_0{>}0$ renders the homogeneous gauge sector $\mathbf{g}_\mathrm{gs}{=}(g^\mathrm{gs}_1,\ldots,g^\mathrm{gs}_L)$ as the ground-state manifold, where $g^\mathrm{gs}_j{=}0,\,\forall j$, in the case of the $\mathrm{U}(1)$ QLM, and $g^\mathrm{gs}_j{=}-1,\,\forall j$, in the case of the $\mathbb{Z}_2$ LGT. We note that this choice of lowest-energy homogeneous gauge sector is not unique. As such, the ground state of $\hat{H}_\mathrm{prep}$ is a domain wall in the matter fields, with the left half of the chain fully occupied, while the right half is empty, residing in the lowest-energy homogeneous gauge sector.

\begin{figure}[t!]
    \centering
    \includegraphics[width=\columnwidth]{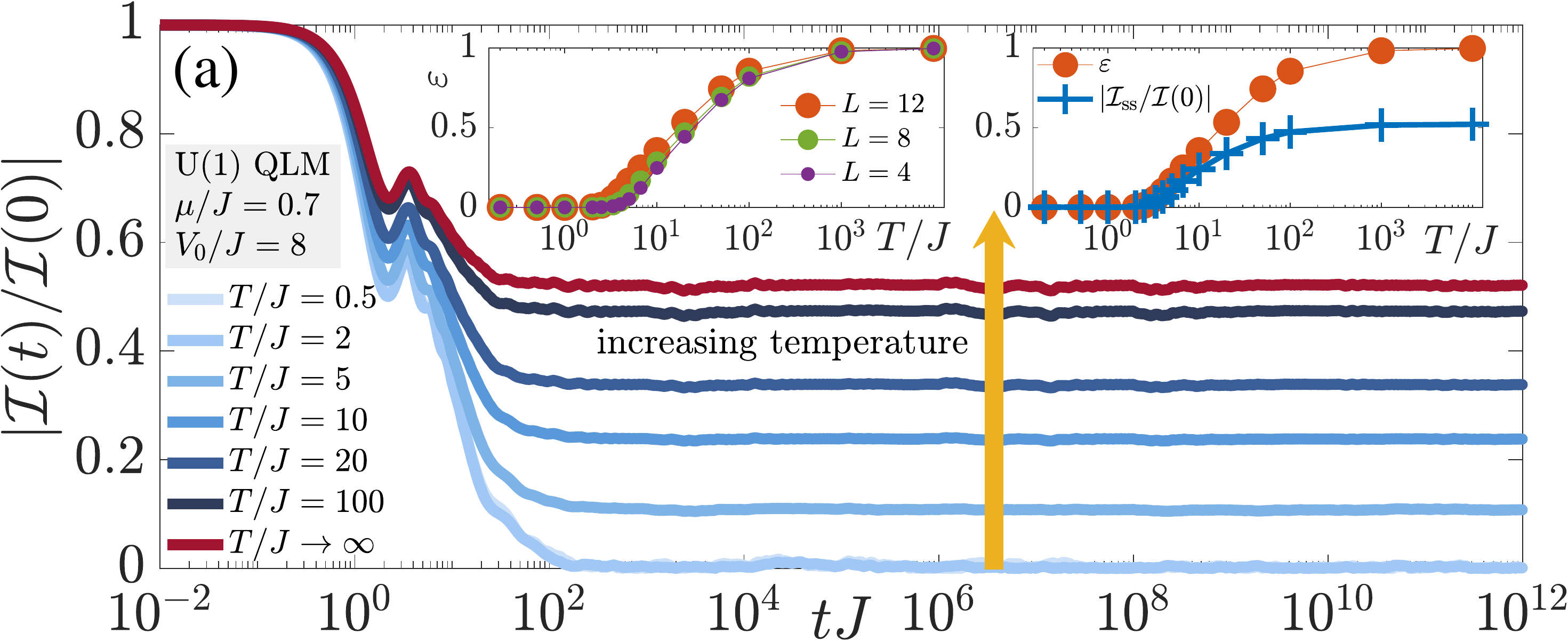}\\
    \vspace{2.1mm}
    \includegraphics[width=\columnwidth]{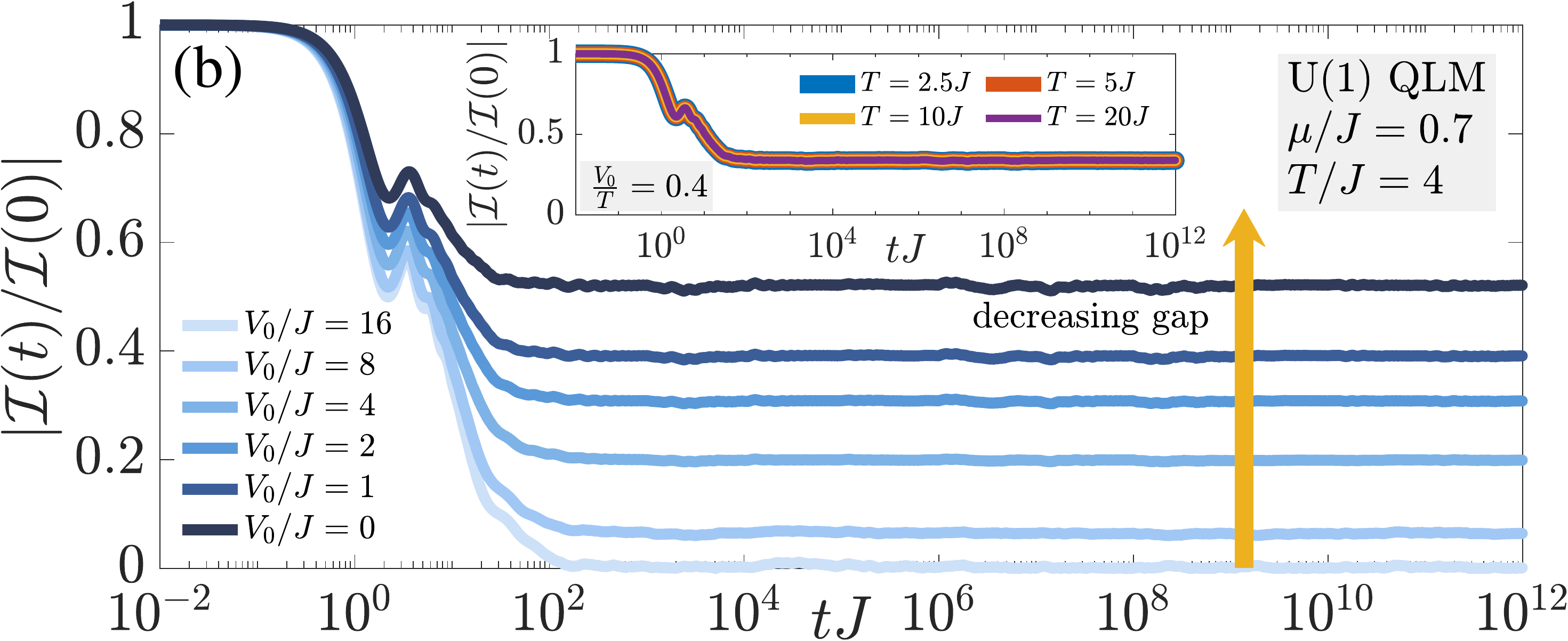}
    \caption{(Color online). Dynamics of the normalized imbalance starting in the canonical thermal ensemble~\eqref{eq:CE} with temperature $T{=}1/\beta$ and quenching with the $\mathrm{U}(1)$ QLM Hamiltonian~\eqref{eq:H0_U1}. Results are computed in exact diagonalization for $L{=}12$ sites unless otherwise specified. (a) The plateau value of the normalized imbalance at long times increases with temperature, indicating greater localization when the system is prepared in hotter initial ensembles. The insets show a direct connection between the value of gauge violation~\eqref{eq:viol} and degree of localization. (b) For a fixed ensemble temperature $T$, a smaller gap $V_0$ in the preparation Hamiltonian~\eqref{eq:prep} leads to greater localization, because it facilitates a greater mixture of gauge sectors. The inset shows how the normalized-imbalance dynamics is identical for values of the temperature and gap at a fixed ratio $V_0{/}T$.}
    \label{fig:U1QLM}
\end{figure}

We now prepare the system in the canonical ensemble ($k_\mathrm{B}{=}1$)
\begin{align}\label{eq:CE}
    \hat{\rho}_0=\frac{e^{-\beta\hat{H}_\mathrm{prep}}}{\mathcal{Z}}=\sum_\mathbf{g}p_\mathbf{g}\hat{\rho}_\mathbf{g},
\end{align}
where $T{=}1/\beta$ is the ensemble temperature, $\hat{\mathcal{P}}_\mathbf{g}$ is the projector onto the gauge sector $\mathbf{g}$, $\hat{\rho}_\mathbf{g}{=}\hat{\mathcal{P}}_\mathbf{g}e^{-\beta\hat{H}_\mathrm{prep}}\hat{\mathcal{P}}_\mathbf{g}/\mathcal{Z}_\mathbf{g}$ and $\mathcal{Z}_\mathbf{g}{=}\Tr\big\{\hat{\mathcal{P}}_\mathbf{g}e^{-\beta\hat{H}_\mathrm{prep}}\big\}$ are the initial density operator and the partition function within sector $\mathbf{g}$, $\mathcal{Z}{=}\sum_\mathbf{g}\mathcal{Z}_\mathbf{g}$ is the total partition function, $p_\mathbf{g}{=}\mathcal{Z}_\mathbf{g}/\mathcal{Z}$ is the sector weight, and we have utilized $\hat{H}_\mathrm{prep}{=}\sum_\mathbf{g}\hat{\mathcal{P}}_\mathbf{g}\hat{H}_\mathrm{prep}\hat{\mathcal{P}}_\mathbf{g}$ since $\big[\hat{H}_\mathrm{prep},\hat{G}_j\big]{=}0,\,\forall j$. 

We now quench $\hat{\rho}_0$ with $\hat{H}_0$ at $t{=}0$, which gives rise to the time-evolved density operator of the system ($\hbar{=}1$)
\begin{align}\label{eq:TimeEvolution}
    \hat{\rho}(t)=e^{-i\hat{H}_0t}\hat{\rho}_0e^{i\hat{H}_0t}=\sum_\mathbf{g}p_\mathbf{g}\hat{\rho}_\mathbf{g}(t),
\end{align}
where $\hat{\rho}_\mathbf{g}(t){=}\hat{\mathcal{P}}_\mathbf{g}e^{-i\hat{H}_0t}\hat{\rho}_\mathbf{g}e^{i\hat{H}_0t}\hat{\mathcal{P}}_\mathbf{g}$, as due to the gauge symmetry of $\hat{H}_0$ we are able to write $\hat{H}_0{=}\sum_\mathbf{g}\hat{\mathcal{P}}_\mathbf{g}\hat{H}_0\hat{\mathcal{P}}_\mathbf{g}$. Equation~\eqref{eq:TimeEvolution} allows for the calculation of the dynamics in each gauge sector separately, thereby allowing us to reach larger system sizes for the desired evolution times in our numerical simulations. 

$T$-DFL is intuitively expected by examining Eq.~\eqref{eq:TimeEvolution}, and noting that the dynamics of an observable $\hat{A}$ will be equal to a weighted sum of its dynamics in all gauge sectors: $\Tr\big\{\hat{A}\hat{\rho}(t)\big\}{=}\sum_\mathbf{g}p_\mathbf{g}\Tr\big\{\hat{A}\hat{\rho}_\mathbf{g}\big\}$. Consequently, the dynamics of the imbalance will encounter an effective disorder over the background charges of an extensive number of gauge sectors at sufficiently high $T$, which in turn gives rise to $T$-DFL. We now present exact diagonalization results demonstrating this.

We are interested in the dynamics of the time-averaged imbalance,
\begin{align}\label{eq:imbalance}
    \mathcal{I}(t)=\frac{1}{Lt}\int_0^tds\sum_{j=1}^L\alpha_j\Tr\big\{\hat{\rho}(s)\hat{n}_j\big\},
\end{align}
as a function of temperature $T{=}1/\beta$. Furthermore, we want to relate the (de)localization to the gauge violation in the canonical ensemble, defined as
\begin{align}\label{eq:viol}
    \varepsilon=\frac{1}{L}\sum_{j=1}^L\Tr\Big\{\hat{\rho}_0\big(\hat{G}_j-g^\mathrm{gs}_j\big)^2\Big\},
\end{align}
which is conserved throughout the quench dynamics.

Figure~\ref{fig:U1QLM} shows the resulting quench dynamics of the imbalance~\eqref{eq:imbalance} normalized by its value at $t{=}0$ for the case of the spin-$1/2$ $\mathrm{U}(1)$ QLM with $L{=}12$ sites. At $t{=}0$ and $T{=}0$, the system is in the ground state of $\hat{H}_\mathrm{prep}$, which is a gauge-invariant domain-wall state in a homogeneous gauge sector, and thus $\varepsilon{=}0$. A quench at zero temperature leads to ergodic behavior where the imbalance, initially $0.5$ at $t{=}0$, decays to zero at long times, indicating thermalization and loss of initial-state memory. This behavior persists at small values of $T$ where $\varepsilon{\approx}0$; see Fig.~\ref{fig:U1QLM}(a) and its insets. When the temperature is large enough such that $\varepsilon$ is appreciably nonzero, we see that the imbalance settles into a finite nonzero plateau that persists for all calculated evolution times, indicating DFL and the absence of thermalization. The value of the plateau increases with temperature and reaches a maximum at asymptotically large $T\to\infty$ $(\beta\to0)$\footnote{Of course, if $T$ is infinite, the imbalance will be zero at all times: $\mathcal{I}(0){=}\mathcal{I}(t){=}0$, and normalized imbalance is not defined. Hence, by $T{\to}\infty$ we denote asymptotically large temperatures and do not mean infinite temperature.}. In the right inset of Fig.~\ref{fig:U1QLM}(a), the steady-state value of the normalized imbalance shows a direct correspondence with the gauge violation, and is only finite when the latter is. We have checked that the steady-state plateau cannot be faithfully described by a thermal ensemble, where the latter predicts a zero long-time imbalance for any value of $T$; see Supplemental Material (SM) \cite{SM}. We emphasize that at no time during the evolution are there any coherences between different gauge sectors. Such coherences can be created by adding a polarizing field ${\propto}\sum_j\hat{s}^x_{j,j+1}$ to $\hat{H}_\mathrm{prep}$, similar to what is done in the traditional case of DFL from a quantum superposition pure initial state \cite{Smith2017,Brenes2018}, but this is not necessary for \textit{temperature-induced} DFL.

The gap $V_0$ of the ground-state sector also affects the degree of localization. The gap $V_0$ in Eq.~\eqref{eq:prep} isolates the homogeneous sector by making it a ground-state manifold. At larger $V_0$, a higher-temperature ensemble is required to achieve localization, since the density of states is negligible at low energies $E_0{-}E_\text{gs}{\lesssim}V_0$ above the ground state; see Fig.~\ref{fig:schematic}. Equivalently, the smaller $V_0$ is at a given temperature, the more likely is the system to localize, as shown in Fig.~\ref{fig:U1QLM}(b). Indeed, in the case of $V_0{=}0$, DFL can already occur at $T{\approx}0$. In that case, there is no unique lowest-lying gauge sector, and the ground-state manifold of $\hat{H}_\mathrm{prep}$ will consist of $2^L$ degenerate domain-wall states residing in an extensive number of gauge sectors. In the inset of Fig.~\ref{fig:U1QLM}(b), we show the intimate connection between the gap $V_0$ and the ensemble temperature $T$, where the DFL behavior is identical at different values of these parameters at a fixed value of $V_0{/}T$. Even though we have chosen $S{=}1/2$ and $\mu{=}0.7J$ for the numerical simulations in Fig.~\ref{fig:U1QLM}, we have checked that our conclusions remain the same for different values of link spin and fermionic mass \cite{SM}.

\begin{figure}[t!]
    \centering
    \includegraphics[width=\columnwidth]{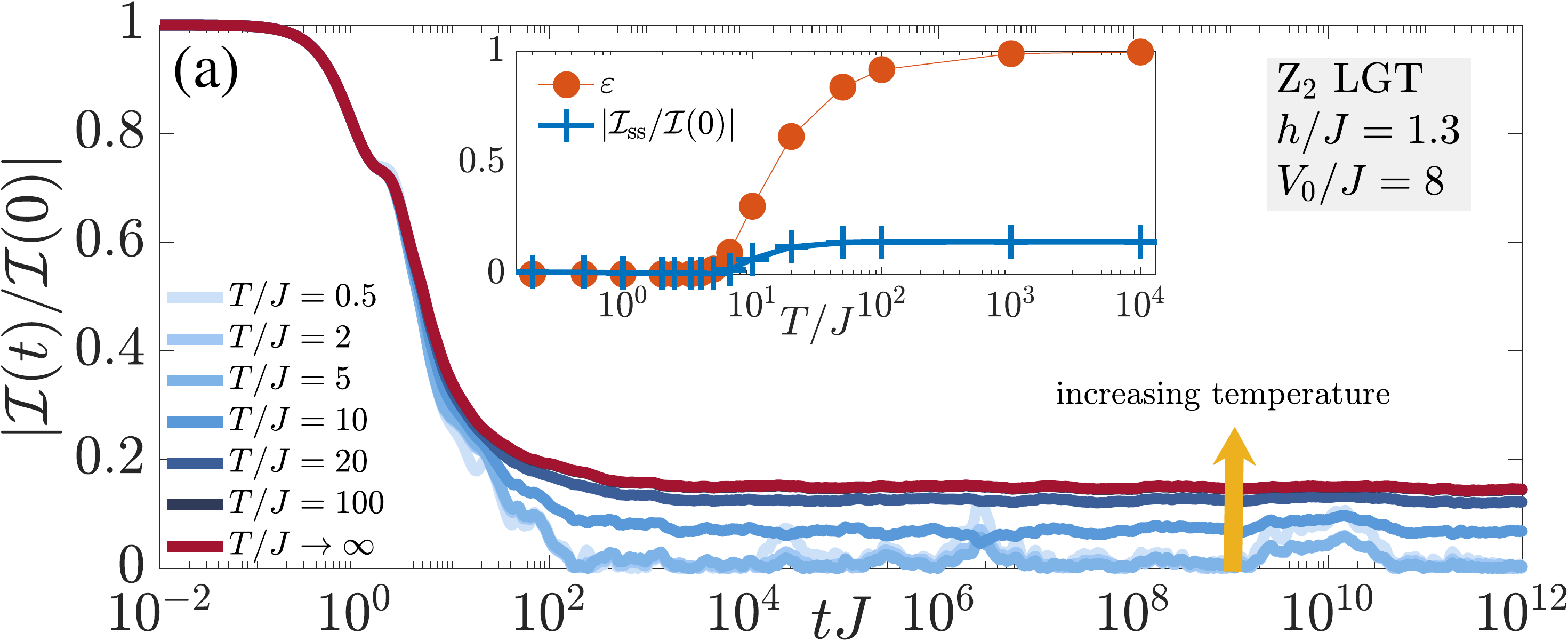}\\
    \vspace{2.1mm}
    \includegraphics[width=\columnwidth]{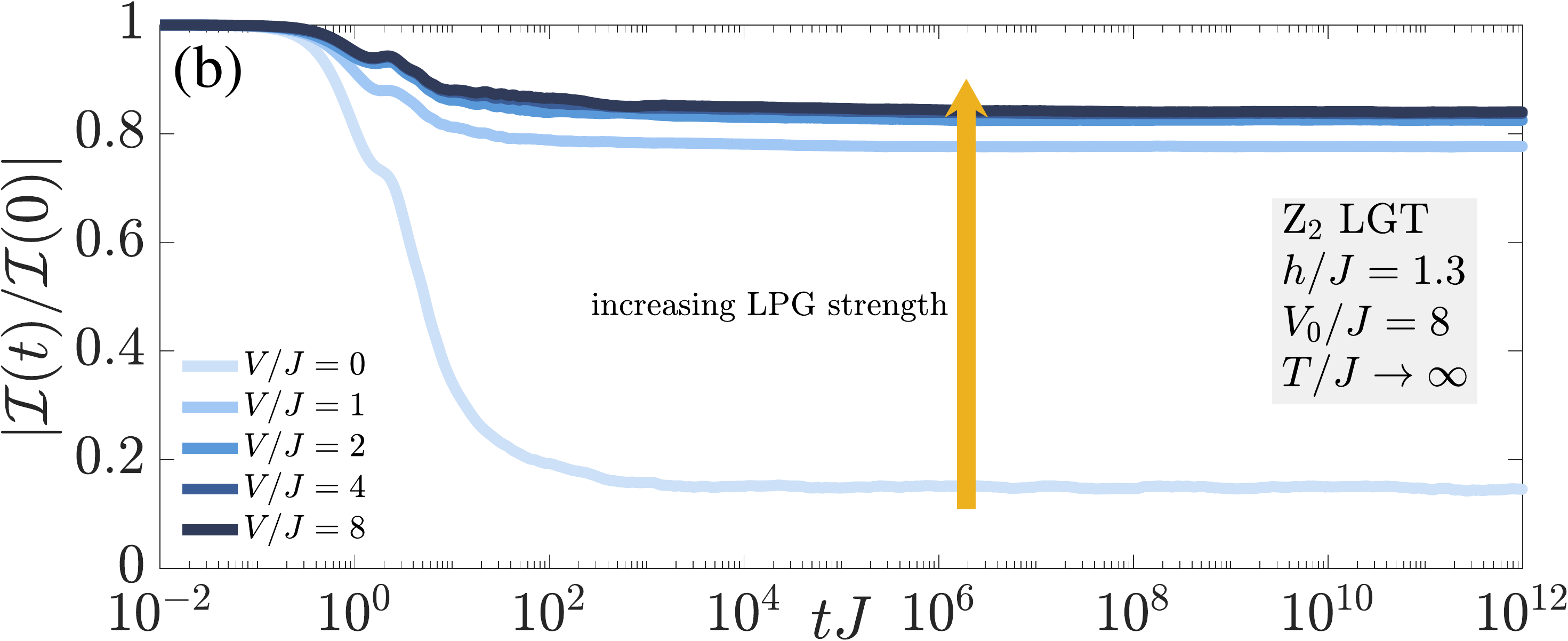}
    \caption{(Color online). Dynamics of the normalized imbalance starting in the canonical thermal ensemble~\eqref{eq:CE} with temperature $T{=}1/\beta$ and quenching with the $\mathbb{Z}_2$ LGT Hamiltonian~\eqref{eq:H0_Z2}. Results are computed in exact diagonalization for $L{=}8$ sites. (a) The normalized imbalance settles at long times into a plateau whose value increases with temperature. (b) Adding a term linear in the local pseudogenerators~\eqref{eq:LPG} to the quench Hamiltonian leads to an enhancement of disorder-free localization for a fixed temperature.}
    \label{fig:Z2LGT}
\end{figure}

Next, we look at $T$-DFL in the $\mathbb{Z}_2$ LGT~\eqref{eq:H0_Z2} in Fig.~\ref{fig:Z2LGT}. The picture is qualitatively identical to that of the $\mathrm{U}(1)$ QLM in Fig.~\ref{fig:U1QLM}(a), where we see a direct connection between the ensemble temperature $T$ and the steady-state value of the normalized-imbalance plateau in Fig.~\ref{fig:Z2LGT}(a). The larger $T$ is, the more prominent DFL is, with a larger value for long-time normalized-imbalance plateau. We also checked that a thermal ensemble does not faithfully describe this plateau, predicting instead a zero imbalance in the long-time limit regardless of the value of $T$. As in the case of the $\mathrm{U}(1)$ QLM, the steady-state value of the normalized imbalance shows a direct monotonic relation with the gauge violation, see inset of Fig.~\ref{fig:Z2LGT}(a), indicating that $T$-DFL occurs only when $\varepsilon>0$ in the initial thermal ensemble.

It is worth noting that the $T$-DFL is markedly weaker for the $\mathbb{Z}_2$ LGT compared to the $\mathrm{U}(1)$ QLM. This is in agreement with results on DFL starting in a superposition initial state \cite{Halimeh2021stabilizingDFL,Halimeh2021enhancing}. The origin of this difference is that the $\mathbb{Z}_2$ LGT can locally admit only two charges ${\pm}1$ due to the underlying $\mathbb{Z}_2$ gauge symmetry, thus leading to a restricted form of discrete binary disorder. In the case of the $\mathrm{U}(1)$ QLM with a spin-$1/2$ representation, a local constraint admits four different charges $g_j{=}(-1)^jq_j$, with $q_j{\in}\{-1,0,1,2\}$. This naturally leads to a wider variety of gauge sectors, which then allows for a greater effective disorder over their background charges in a thermal ensemble. However, one can enhance the symmetry of the $\mathbb{Z}_2$ LGT by introducing the term $V\hat{H}_W{=}V\sum_jj\hat{W}_j$ where the local pseudogenerator (LPG) is defined as \cite{Halimeh2021stabilizing}
\begin{align}\label{eq:LPG}
    \hat{W}_j=\hat{\tau}^x_{j-1,j}\hat{\tau}^x_{j,j+1}+2g^\mathrm{gs}_j\hat{n}_j.
\end{align}
The LPG is identical to the local full generator $\hat{G}_j^{\mathbb{Z}_2}$ in the homogeneous sector $\mathbf{g}_\text{gs}$: $\hat{G}_j^{\mathbb{Z}_2}\ket{\phi}{=}g^\mathrm{gs}_j\ket{\phi}{\iff}\hat{W}_j\ket{\phi}{=}g^\mathrm{gs}_j\ket{\phi}$. It was initially developed as an experimentally feasible scheme for stabilizing $\mathbb{Z}_2$ LGTs \cite{Halimeh2021stabilizing,Homeier2022quantum}, but has recently been shown to lead to an effective Hamiltonian that enhances DFL when starting in a superposition initial state \cite{Halimeh2021enhancing,Lang2022stark}. This effective Hamiltonian hosts an enriched local symmetry associated with $\hat{W}_j$ that contains the $\mathbb{Z}_2$ gauge symmetry of Eq.~\eqref{eq:H0_Z2}. This means that now there are more local-symmetry superselection sectors over which the effective disorder can be generated, and this leads to more prominent DFL in the case of a superposition initial state \cite{Halimeh2021enhancing}. It would be interesting to see if this enhancement also occurs in the case of a system prepared in a thermal ensemble. For this purpose, we consider a canonical ensemble with an asymptotically large temperature $T{\to}\infty\,(\beta{\to}0)$ and quench it with $\hat{H}_0^{\mathbb{Z}_2}{+}V\hat{H}_W$. The ensuing imbalance dynamics is shown in Fig.~\ref{fig:Z2LGT}(b). As the LPG-term strength $V$ is increased, the DFL is drastically enhanced. It is important to note here that the addition of the LPG term does not alter the value of $\varepsilon$ because $\big[\hat{G}_j^{\mathbb{Z}_2},\hat{W}_j\big]{=}0$. The enhancement of DFL occurs strictly from the local symmetry that emerges due to the LPG term. This behavior is also present at finite temperatures, where if DFL is present at a given temperature $T$ for $V{=}0$, it will get enhanced for $V{>}0$. In case the system initialized at temperature $T$ is delocalized for $V{=}0$, it will remain delocalized for $V{>}0$.

\textbf{\textit{Discussion and outlook.---}}We have demonstrated temperature-induced disorder-free localization, where a LGT initialized in a thermal canonical ensemble at sufficiently high temperature will exhibit DFL in its quench dynamics up to all accessible evolution times. No polarizing field is employed to create any kind of coherent superposition over the gauge sectors, and the preparation Hamiltonian is chosen such that at zero temperature the system will reside only in a homogeneous gauge sector where no DFL is possible.

We have illustrated our findings in two paradigmatic systems: the $\mathrm{U}(1)$ quantum link formulation of the Schwinger model and the $\mathbb{Z}_2$ lattice gauge theory. The qualitative picture in both is the same: the larger the temperature of the initial ensemble, the more localized will the dynamics of the system be at late times. Additionally, we have shown how the gap of the preparation Hamiltonian can also influence DFL, with smaller gaps leading to more prominent DFL at a given temperature. A given ratio of the gap and ensemble temperature will always lead to the same degree of localization. In the case of the $\mathbb{Z}_2$ LGT, linear weighted sums in the local pseudogenerator can be employed to dynamically induce an emergent enriched local symmetry that leads to enhanced $T$-DFL. Furthermore, the choice of the preparation Hamiltonian is not unique to Eq.~\eqref{eq:prep}. In fact, for the $\mathbb{Z}_2$ LGT one can replace the gap term with $V_0\sum_j\hat{W}_j$, which offers greater experimental feasibility and leads to $T$-DFL, although the temperature-dependence may become different. It is also worth noting that $T$-DFL can be stabilized against gauge-breaking errors using linear gauge protection \cite{Halimeh2021stabilizingDFL,Halimeh2021enhancing,SM}.

Intriguingly, even though DFL is a quantum interference phenomenon, here we show it gets more pronounced at high temperatures. Hence, one might expect that in contrast to common intuition that low-temperature physics probes quantum effects, there might be other phenomena of quantum nonergodicity becoming prominent at high temperature.

Our findings can be tested in ultracold-atom platforms \cite{Goerg2019,Schweizer2019,Mil2020,Yang2020,Zhou2021} and on digital quantum computers \cite{Arute2019}, which recently have become a viable framework for simulating various gauge-theory phenomena \cite{Martinez2016,Wang2022,Mildenberger2022}. One way to generate a thermal ensemble on a quantum computer is via a purification  called a thermofield double state \cite{Wu2019,Sagastizabal2021,Zhang2021CV}. The latter can be constructed on two identical subsystems of the circuit in an enlarged Hilbert space. The desired Gibbs ensemble on one subsystem is then realized by tracing out the second subsystem.

Our numerical results indicate that $T$-DFL in our $1{+}1$D setting does not include a finite-temperature delocalization-localization transition but only a crossover. It would be interesting to explore such a possible transition in $2{+}1$D, e.g., in Rydberg tweezer arrays \cite{Homeier2022quantum}.

\bigskip
\begin{acknowledgments}
J.C.H.~is grateful to Haifeng Lang, Achilleas Lazarides, Pablo Sala, and Guo-Xian Su for stimulating discussions. J.C.H.~and F.G.~acknowledge funding from the European Research Council (ERC) under the European Union’s Horizon 2020 research and innovation programm (Grant Agreement no 948141) — ERC Starting Grant SimUcQuam, and by the Deutsche Forschungsgemeinschaft (DFG, German Research Foundation) under Germany's Excellence Strategy -- EXC-2111 -- 390814868. P.H.~acknowledges support by the ERC Starting Grant StrEnQTh (project ID 804305), the Google Research Scholar Award ProGauge, Provincia Autonoma di Trento, and Q@TN — Quantum Science and Technology in Trento. J.K.~acknowledges support from the Imperial--TUM flagship partnership. The research is part of the Munich Quantum Valley, which is supported by the Bavarian state government with funds from the Hightech Agenda Bayern Plus.
\end{acknowledgments}

\clearpage
\pagebreak
\newpage
\setcounter{equation}{0}
\setcounter{figure}{0}
\setcounter{table}{0}
\setcounter{page}{1}
\makeatletter
\renewcommand{\bibnumfmt}[1]{[S#1]}
\renewcommand{\citenumfont}[1]{S#1}
\renewcommand{\theequation}{S\arabic{equation}}
\renewcommand{\thefigure}{S\arabic{figure}}
\renewcommand{\thetable}{S\Roman{table}}
\renewcommand{\bibnumfmt}[1]{[S#1]}
\renewcommand{\citenumfont}[1]{S#1}
\widetext
\begin{center}
\textbf{--- Supplemental Material ---\\Temperature-Induced Disorder-Free Localization}
\text{Jad C.~Halimeh, Philipp Hauke, Johannes Knolle, and Fabian Grusdt}
\end{center}

\date{\today}
\maketitle

\section{Nonthermal DFL steady state}
Here we show that the steady state reached in the imbalance at finite temperature cannot be described by a thermal ensemble, even though the system is initially in a thermal ensemble at $t=0$. Let us assume by way of contradiction that the late-time steady state can be described by the thermal Gibbs ensemble
\begin{align}
    \hat{\rho}^\text{CE}_\mathrm{ss}=\frac{e^{-\beta_\mathrm{ss}\hat{H}}}{\Tr\big\{e^{-\beta_\mathrm{ss}\hat{H}}\big\}},
\end{align}
where $T_\mathrm{ss}=1/\beta_\mathrm{ss}$ is its temperature, and $\hat{H}$ is the quench Hamiltonian, which in our case is either $\hat{H}_0$ or $\hat{H}_0^{\mathbb{Z}_2}+V\hat{H}_W$. The global quenches considered in our work are unitary, and thus the quench energy must be conserved at all times:
\begin{align}\label{eq:energy}
    E_\text{quench}=\Tr\big\{\hat{H}\hat{\rho}_0\big\}\mbeq\Tr\big\{\hat{H}\hat{\rho}^\text{CE}_\mathrm{ss}\big\}.
\end{align}
Equation~\eqref{eq:energy} is now an implicit equation with the only unknown being $\beta_\mathrm{ss}$, which can be solved using, e.g., Newton's method. Hence, $\hat{\rho}^\text{CE}_\mathrm{ss}$ can be numerically determined, and one can check whether $\Tr\big\{\hat{A}\hat{\rho}^\text{CE}_\mathrm{ss}\big\}\approx\overline{\Tr\big\{\hat{A}\hat{\rho}(t\to\infty)\big\}}$, for any local observable $\hat{A}$ \cite{Rigol_2008-S}. In case this condition is satisfied, then the system will have thermalized \cite{Deutsch1991-S,Srednicki1994-S}, and otherwise it means the dynamics is nonergodic \cite{Rigol_review-S}. We numerically find that $\hat{\rho}^\text{CE}_\mathrm{ss}$ only predicts a zero imbalance for all the quenches considered in our work, which is in contradiction to the exact dynamics for sufficiently large temperatures when $T$-DFL is present. As such, we conclude that the steady state arising at long times is nonthermal when there is $T$-DFL.

When it comes to the steady state, one can alternatively look at the prediction from a microcanonical ensemble. The latter is constructed from the eigenstates $\ket{\epsilon_m}$ of the quench Hamiltonian that lie within the energy shell $\mathcal{S}=\big[E_\text{quench}-\delta \epsilon,E_\text{quench}+\delta \epsilon\big]$, where $E_\text{quench}=\Tr\big\{\hat{H}\hat{\rho}_0\big\}$ is the quench energy. The microcanonical ensemble then takes the form \cite{Rigol_2008-S}
\begin{align}
    \hat{\rho}^\mathrm{ME}_\text{ss}=\mathcal{N}_{E_\text{quench},\delta \epsilon}^{-1}\sum_{m;\,\epsilon_m\in\mathcal{S}}\ketbra{\epsilon_m}{\epsilon_m},
\end{align}
where $\mathcal{N}_{E_\text{quench},\delta \epsilon}$ is the number of quench-Hamiltonian eigenstates with eigenenergies within the shell $\mathcal{S}$. In our numerical calculations, we have set $\delta E=0.1J$, though we have checked that our qualitative conclusions do not strongly depend on this choice. In our numerical simulations, $\hat{\rho}^\mathrm{ME}_\text{ss}$ also predicts a zero imbalance at long times for all quenches considered, even when $T$-DFL is prominent. This further shows that the finite plateau is described by a nonthermal steady state.

\section{Temperature-induced DFL in the spin-$S$ $\mathrm{U}(1)$ QLM with $S>1/2$}
In the main text, our analysis of the $\mathrm{U}(1)$ QLM has focused on a spin-$1/2$ representation of the local gauge fields. However, $T$-DFL is not restricted to that case, and higher-$S$ representations will also exhibit it.

\begin{figure}[htp]
    \centering
    \includegraphics[width=0.49\columnwidth]{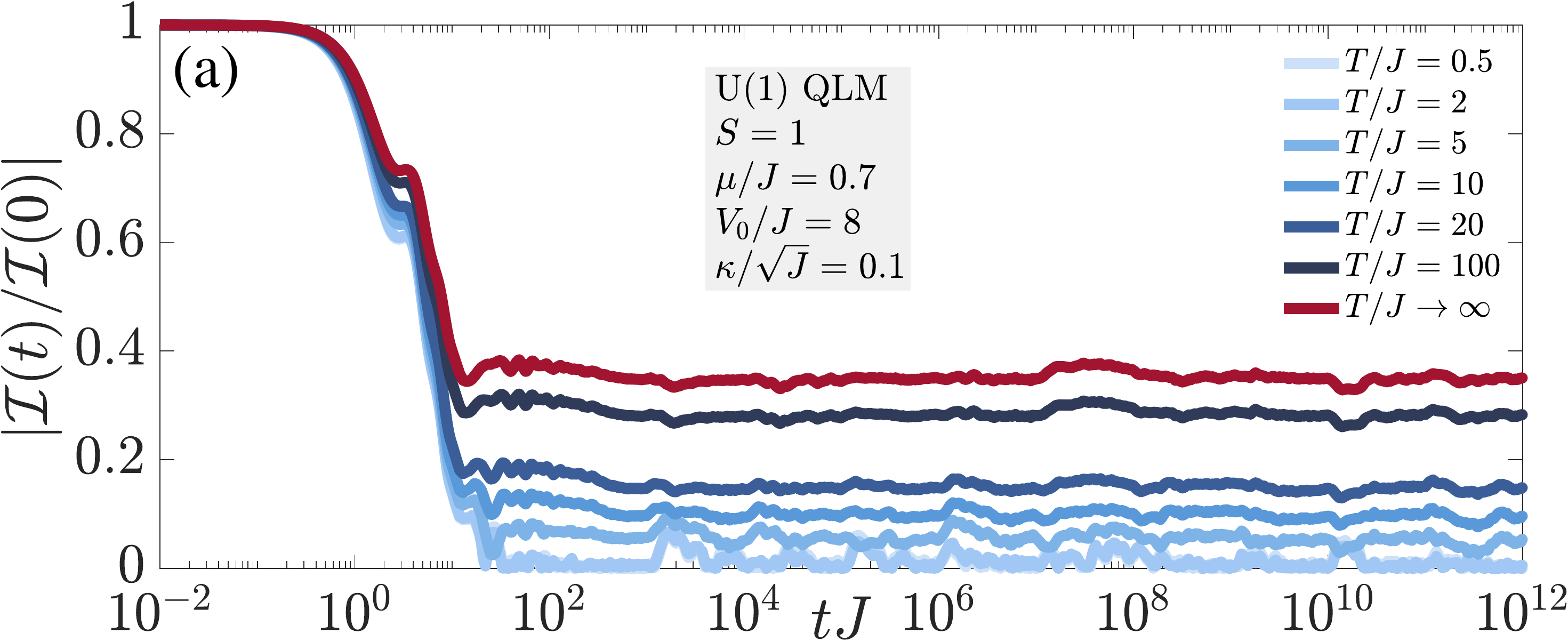}\quad
    \includegraphics[width=0.49\columnwidth]{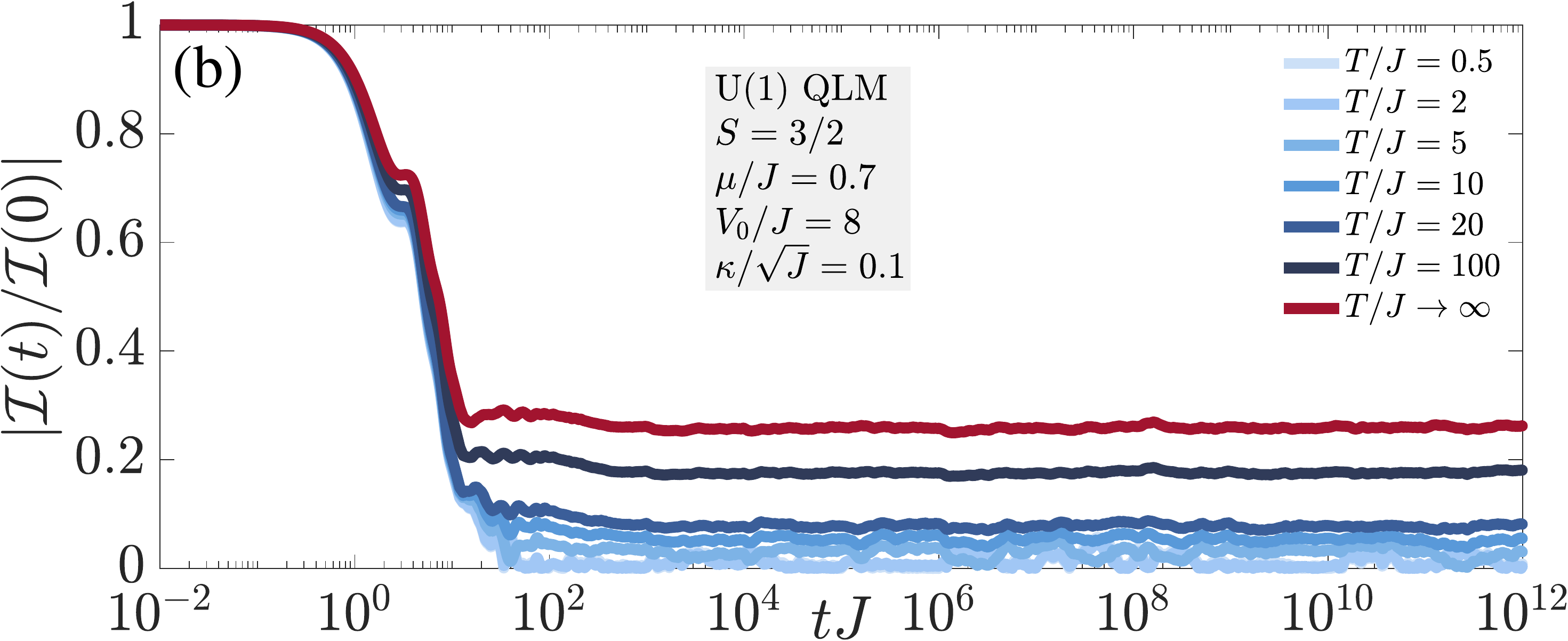}
    \caption{(Color online). Temperature-induced DFL in the (a) spin-$1$ and (b) spin-$3/2$ $\mathrm{U}(1)$ QLM. The qualitative behavior is the same as in the case of the spin-$1/2$ $\mathrm{U}(1)$ QLM, with DFL getting enhanced with initial-ensemble temperature $T$.}
    \label{fig:HigherS}
\end{figure}

To demonstrate this, we show the quench dynamics of the normalized imbalance for the spin-$1$ and spin-$3/2$ $\mathrm{U}(1)$ QLM in Fig.~\ref{fig:HigherS}(a,b), respectively, for $L=4$ matter sites, $\mu=0.7J$ and $\kappa=0.1\sqrt{J}$. The qualitative picture is the same as the spin-$1/2$ case, where $T$-DFL arises at sufficiently high ensemble temperatures, and the degree of localization displays a monotonic relationship with the ensemble temperature. We have also checked that the finite plateau arising in case of $T$-DFL is not described by a thermal ensemble, where the latter always predicts a zero imbalance at late times. As in the case of the spin-$1/2$ $\mathrm{U}(1)$ QLM, we have checked that the specific values of $\mu$ and $\kappa$ do not alter the qualitative picture of $T$-DFL.

\section{Stability of temperature-induced DFL}
It has been shown that DFL arising from a superposition initial state is unstable in the presence of gauge-breaking perturbations \cite{Smith2018-S}. However, there has recently been experimentally feasible proposals based on linear gauge protection \cite{Halimeh2020e-S,Halimeh2021stabilizing-S} that stabilize and even enhance DFL \cite{Halimeh2021stabilizingDFL-S,Halimeh2021enhancing-S,Lang2022stark-S}. To put things on a formal fitting, let us consider the gauge-breaking terms
\begin{subequations}\label{eq:errors}
\begin{align}
    &\lambda\hat{H}_1^{\mathrm{U}(1)}=\lambda\sum_{j=1}^L\bigg[\hat{\sigma}^+_j\hat{\sigma}^+_{j+1}+\hat{\sigma}^-_j\hat{\sigma}^-_{j+1}+\frac{\hat{s}^x_{j,j+1}}{2\sqrt{S(S+1)}}\bigg],\\
    &\lambda\hat{H}_1^{\mathbb{Z}_2}=\lambda\sum_{j=1}^L\big[\hat{a}_j^\dagger\hat{a}_{j+1}+\hat{a}_j\hat{a}_{j+1}^\dagger+\hat{\tau}z_{j,j+1}\big],
\end{align}
\end{subequations}
at strength $\lambda$, relevant to synthetic quantum matter implementations of $\mathrm{U}(1)$ QLMs and $\mathbb{Z}_2$ LGTs with both dynamical matter and gauge fields \cite{Schweizer2019-S,Mil2020-S,Yang2020-S,Zhou2021-S}. These error terms involve tunneling of matter without a concomitant change in the electric field to preserve Gauss's law, or vice versa. To protect against them, we add the terms \cite{Halimeh2020e-S,Halimeh2021stabilizing-S,Lang2022stark-S}
\begin{subequations}
\begin{align}\label{eq:LinProU1}
    &V\hat{H}_G=V\sum_{j=1}^Lj\hat{G}^{\mathrm{U}(1)}_j,\\\label{eq:LinProZ2}
    &V\hat{H}_W=V\sum_{j=1}^Lj\hat{W}_j,
\end{align}
\end{subequations}
where the full local generator of the $\mathrm{U}(1)$ gauge symmetry is given in Eq.~\eqref{eq:Gj_U1} and the local pseudogenerator $\hat{W}_j$ is defined in Eq.~\eqref{eq:LPG} for the case of the $\mathbb{Z}_2$ LGT.

\begin{figure}[htp]
    \centering
    \includegraphics[width=0.49\columnwidth]{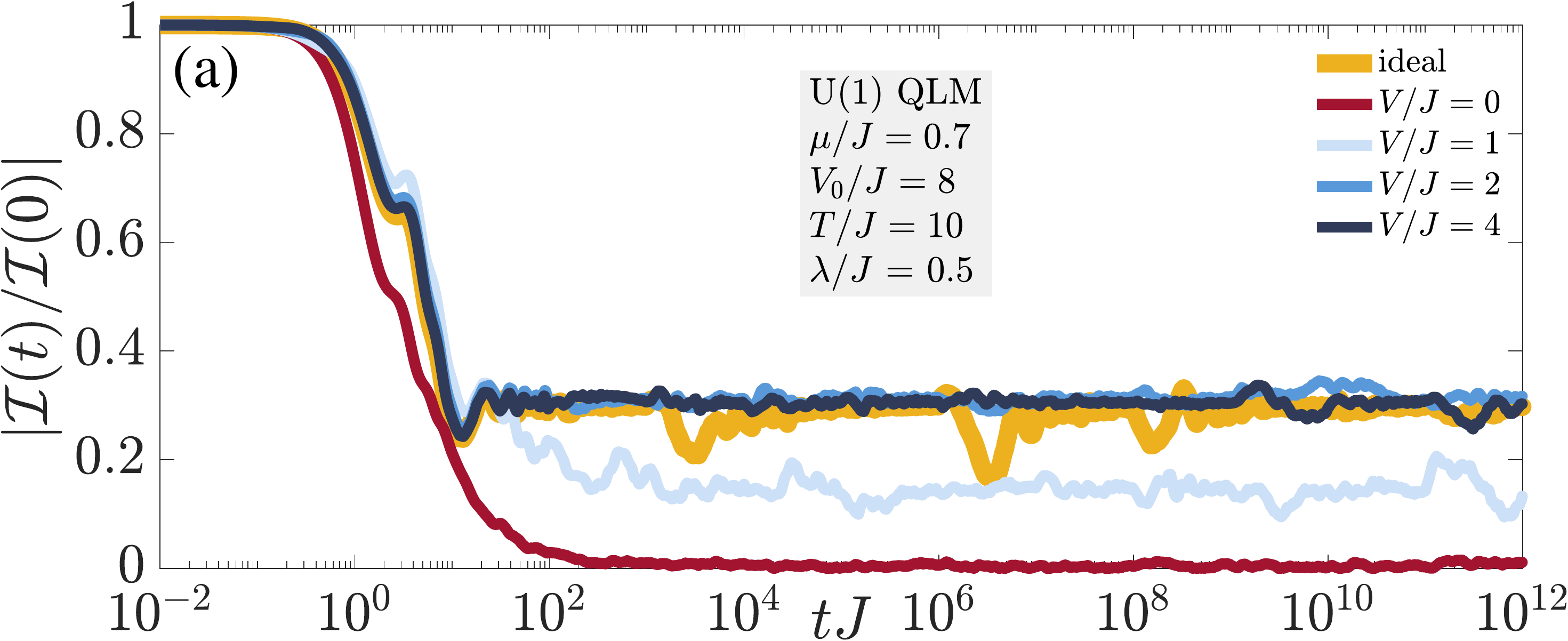}\quad
    \includegraphics[width=0.49\columnwidth]{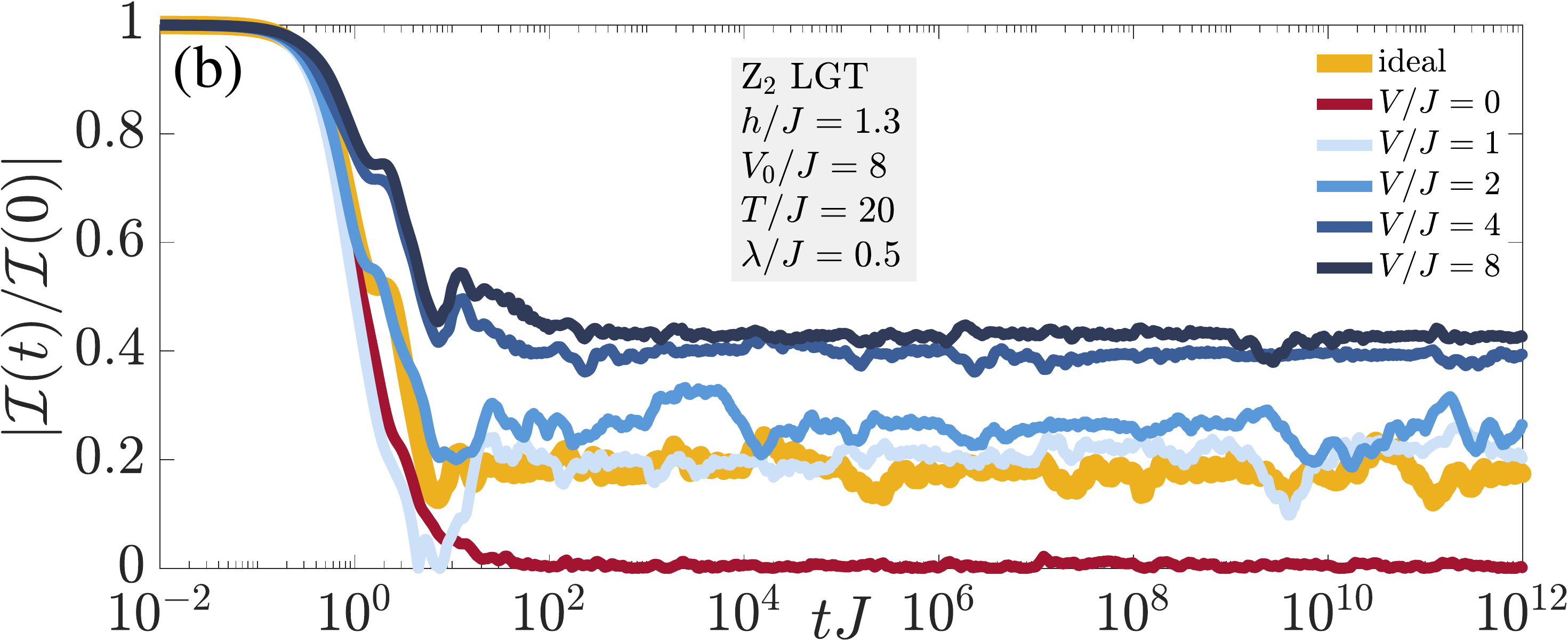}
    \caption{(Color online). Linear gauge protection against gauge-breaking errors in (a) the spin-$1/2$ $\mathrm{U}(1)$ QLM at $T=10J$ using Eq.~\eqref{eq:LinProU1}, and (b) the $\mathbb{Z}_2$ LGT at $T=20J$ using Eq.~\eqref{eq:LinProZ2}. In both cases we see that without protection, the gauge-breaking errors~\eqref{eq:errors} completely destroy $T$-DFL. At sufficiently large $V$, however, $T$-DFL is restored, and even enhanced in the case of the $\mathbb{Z}_2$ LGT due to an emergent enriched local symmetry associated with the LPGs $\hat{W}_j$, defined in Eq.~\eqref{eq:LPG}.}
    \label{fig:protection}
\end{figure}

We now quench our initial thermal ensemble with $\hat{H}=\hat{H}_0+\lambda\hat{H}_1+V\hat{H}_\mathrm{pro}$, where $\hat{H}_\mathrm{pro}=\hat{H}_G$ for the case of the $\mathrm{U}(1)$ QLM and $\hat{H}_\mathrm{pro}=\hat{H}_W$ for the case of the $\mathbb{Z}_2$ LGT. Due to the gauge-breaking term $\lambda\hat{H}_1$, the time-evolved density operator $\hat{\rho}(t)=e^{-i\hat{H}t}\hat{\rho}_0e^{i\hat{H}t}$ cannot be written in the form~\eqref{eq:TimeEvolution}, as now $\hat{H}$ includes gauge-breaking terms $\lambda\hat{H}_1$ that couple different gauge sectors. This renders the numerical simulations more tasking, and as such we restrict our results to $L=4$ matter sites. As shown in Fig.~\ref{fig:protection}(a) for the case of the $\mathrm{U}(1)$ QLM, a finite value of $\lambda$ will completely destroy $T$-DFL in the absence of protection ($V=0$), where the imbalance will go to zero in agreement with a thermal-ensemble prediction (see first section). However, upon turning on linear gauge protection ($V>0$), we see a stabilization of the imbalance, where the temperature-induced DFL is qualitatively restored, with greater quantitative agreement with the ideal case the greater $V$ is.

Similarly in the case of the $\mathbb{Z}_2$ LGT, shown in Fig.~\ref{fig:protection}(b), unprotected gauge-breaking errors destroy DFL. Upon adding the LPG term~\eqref{eq:LinProZ2}, however, the DFL is restored and also enhanced. This stems from the dynamical emergence of an enriched local symmetry due to the LPGs $\hat{W}_j$. This enriched local symmetry generates a greater number of local-symmetry sectors, leading to a greater effective disorder in the thermal average of the imbalance, and hence enhanced localization \cite{Halimeh2021enhancing-S,Lang2022stark-S}.


\begin{thebibliography}{63}%
\makeatletter
\providecommand \@ifxundefined [1]{%
 \@ifx{#1\undefined}
}%
\providecommand \@ifnum [1]{%
 \ifnum #1\expandafter \@firstoftwo
 \else \expandafter \@secondoftwo
 \fi
}%
\providecommand \@ifx [1]{%
 \ifx #1\expandafter \@firstoftwo
 \else \expandafter \@secondoftwo
 \fi
}%
\providecommand \natexlab [1]{#1}%
\providecommand \enquote  [1]{``#1''}%
\providecommand \bibnamefont  [1]{#1}%
\providecommand \bibfnamefont [1]{#1}%
\providecommand \citenamefont [1]{#1}%
\providecommand \href@noop [0]{\@secondoftwo}%
\providecommand \href [0]{\begingroup \@sanitize@url \@href}%
\providecommand \@href[1]{\@@startlink{#1}\@@href}%
\providecommand \@@href[1]{\endgroup#1\@@endlink}%
\providecommand \@sanitize@url [0]{\catcode `\\12\catcode `\$12\catcode
  `\&12\catcode `\#12\catcode `\^12\catcode `\_12\catcode `\%12\relax}%
\providecommand \@@startlink[1]{}%
\providecommand \@@endlink[0]{}%
\providecommand \url  [0]{\begingroup\@sanitize@url \@url }%
\providecommand \@url [1]{\endgroup\@href {#1}{\urlprefix }}%
\providecommand \urlprefix  [0]{URL }%
\providecommand \Eprint [0]{\href }%
\providecommand \doibase [0]{http://dx.doi.org/}%
\providecommand \selectlanguage [0]{\@gobble}%
\providecommand \bibinfo  [0]{\@secondoftwo}%
\providecommand \bibfield  [0]{\@secondoftwo}%
\providecommand \translation [1]{[#1]}%
\providecommand \BibitemOpen [0]{}%
\providecommand \bibitemStop [0]{}%
\providecommand \bibitemNoStop [0]{.\EOS\space}%
\providecommand \EOS [0]{\spacefactor3000\relax}%
\providecommand \BibitemShut  [1]{\csname bibitem#1\endcsname}%
\let\auto@bib@innerbib\@empty
\bibitem [{\citenamefont {Basko}\ \emph {et~al.}(2006)\citenamefont {Basko},
  \citenamefont {Aleiner},\ and\ \citenamefont {Altshuler}}]{Basko2006}%
  \BibitemOpen
  \bibfield  {author} {\bibinfo {author} {\bibfnamefont {D.M.}\ \bibnamefont
  {Basko}}, \bibinfo {author} {\bibfnamefont {I.L.}\ \bibnamefont {Aleiner}}, \
  and\ \bibinfo {author} {\bibfnamefont {B.L.}\ \bibnamefont {Altshuler}},\
  }\bibfield  {title} {\enquote {\bibinfo {title} {Metal–insulator transition
  in a weakly interacting many-electron system with localized single-particle
  states},}\ }\href {\doibase https://doi.org/10.1016/j.aop.2005.11.014}
  {\bibfield  {journal} {\bibinfo  {journal} {Annals of Physics}\ }\textbf
  {\bibinfo {volume} {321}},\ \bibinfo {pages} {1126--1205} (\bibinfo {year}
  {2006})}\BibitemShut {NoStop}%
\bibitem [{\citenamefont {Alet}\ and\ \citenamefont
  {Laflorencie}(2018)}]{Alet_review}%
  \BibitemOpen
  \bibfield  {author} {\bibinfo {author} {\bibfnamefont {Fabien}\ \bibnamefont
  {Alet}}\ and\ \bibinfo {author} {\bibfnamefont {Nicolas}\ \bibnamefont
  {Laflorencie}},\ }\bibfield  {title} {\enquote {\bibinfo {title} {Many-body
  localization: An introduction and selected topics},}\ }\href {\doibase
  https://doi.org/10.1016/j.crhy.2018.03.003} {\bibfield  {journal} {\bibinfo
  {journal} {Comptes Rendus Physique}\ }\textbf {\bibinfo {volume} {19}},\
  \bibinfo {pages} {498--525} (\bibinfo {year} {2018})},\ \bibinfo {note}
  {quantum simulation / Simulation quantique}\BibitemShut {NoStop}%
\bibitem [{\citenamefont {Abanin}\ \emph {et~al.}(2019)\citenamefont {Abanin},
  \citenamefont {Altman}, \citenamefont {Bloch},\ and\ \citenamefont
  {Serbyn}}]{Abanin_review}%
  \BibitemOpen
  \bibfield  {author} {\bibinfo {author} {\bibfnamefont {Dmitry~A.}\
  \bibnamefont {Abanin}}, \bibinfo {author} {\bibfnamefont {Ehud}\ \bibnamefont
  {Altman}}, \bibinfo {author} {\bibfnamefont {Immanuel}\ \bibnamefont
  {Bloch}}, \ and\ \bibinfo {author} {\bibfnamefont {Maksym}\ \bibnamefont
  {Serbyn}},\ }\bibfield  {title} {\enquote {\bibinfo {title} {Colloquium:
  Many-body localization, thermalization, and entanglement},}\ }\href {\doibase
  10.1103/RevModPhys.91.021001} {\bibfield  {journal} {\bibinfo  {journal}
  {Rev. Mod. Phys.}\ }\textbf {\bibinfo {volume} {91}},\ \bibinfo {pages}
  {021001} (\bibinfo {year} {2019})}\BibitemShut {NoStop}%
\bibitem [{\citenamefont {Moudgalya}\ \emph {et~al.}(2018)\citenamefont
  {Moudgalya}, \citenamefont {Rachel}, \citenamefont {Bernevig},\ and\
  \citenamefont {Regnault}}]{Moudgalya2018}%
  \BibitemOpen
  \bibfield  {author} {\bibinfo {author} {\bibfnamefont {Sanjay}\ \bibnamefont
  {Moudgalya}}, \bibinfo {author} {\bibfnamefont {Stephan}\ \bibnamefont
  {Rachel}}, \bibinfo {author} {\bibfnamefont {B.~Andrei}\ \bibnamefont
  {Bernevig}}, \ and\ \bibinfo {author} {\bibfnamefont {Nicolas}\ \bibnamefont
  {Regnault}},\ }\bibfield  {title} {\enquote {\bibinfo {title} {Exact excited
  states of nonintegrable models},}\ }\href {\doibase
  10.1103/PhysRevB.98.235155} {\bibfield  {journal} {\bibinfo  {journal} {Phys.
  Rev. B}\ }\textbf {\bibinfo {volume} {98}},\ \bibinfo {pages} {235155}
  (\bibinfo {year} {2018})}\BibitemShut {NoStop}%
\bibitem [{\citenamefont {Turner}\ \emph {et~al.}(2018)\citenamefont {Turner},
  \citenamefont {Michailidis}, \citenamefont {Abanin}, \citenamefont {Serbyn},\
  and\ \citenamefont {Papi{\'c}}}]{Turner2018}%
  \BibitemOpen
  \bibfield  {author} {\bibinfo {author} {\bibfnamefont {C.~J.}\ \bibnamefont
  {Turner}}, \bibinfo {author} {\bibfnamefont {A.~A.}\ \bibnamefont
  {Michailidis}}, \bibinfo {author} {\bibfnamefont {D.~A.}\ \bibnamefont
  {Abanin}}, \bibinfo {author} {\bibfnamefont {M.}~\bibnamefont {Serbyn}}, \
  and\ \bibinfo {author} {\bibfnamefont {Z.}~\bibnamefont {Papi{\'c}}},\
  }\bibfield  {title} {\enquote {\bibinfo {title} {Weak ergodicity breaking
  from quantum many-body scars},}\ }\href {\doibase 10.1038/s41567-018-0137-5}
  {\bibfield  {journal} {\bibinfo  {journal} {Nature Physics}\ }\textbf
  {\bibinfo {volume} {14}},\ \bibinfo {pages} {745--749} (\bibinfo {year}
  {2018})}\BibitemShut {NoStop}%
\bibitem [{\citenamefont {Sala}\ \emph {et~al.}(2020)\citenamefont {Sala},
  \citenamefont {Rakovszky}, \citenamefont {Verresen}, \citenamefont {Knap},\
  and\ \citenamefont {Pollmann}}]{Sala2020}%
  \BibitemOpen
  \bibfield  {author} {\bibinfo {author} {\bibfnamefont {Pablo}\ \bibnamefont
  {Sala}}, \bibinfo {author} {\bibfnamefont {Tibor}\ \bibnamefont {Rakovszky}},
  \bibinfo {author} {\bibfnamefont {Ruben}\ \bibnamefont {Verresen}}, \bibinfo
  {author} {\bibfnamefont {Michael}\ \bibnamefont {Knap}}, \ and\ \bibinfo
  {author} {\bibfnamefont {Frank}\ \bibnamefont {Pollmann}},\ }\bibfield
  {title} {\enquote {\bibinfo {title} {Ergodicity breaking arising from hilbert
  space fragmentation in dipole-conserving hamiltonians},}\ }\href {\doibase
  10.1103/PhysRevX.10.011047} {\bibfield  {journal} {\bibinfo  {journal} {Phys.
  Rev. X}\ }\textbf {\bibinfo {volume} {10}},\ \bibinfo {pages} {011047}
  (\bibinfo {year} {2020})}\BibitemShut {NoStop}%
\bibitem [{\citenamefont {Khemani}\ \emph {et~al.}(2020)\citenamefont
  {Khemani}, \citenamefont {Hermele},\ and\ \citenamefont
  {Nandkishore}}]{Khemani2020}%
  \BibitemOpen
  \bibfield  {author} {\bibinfo {author} {\bibfnamefont {Vedika}\ \bibnamefont
  {Khemani}}, \bibinfo {author} {\bibfnamefont {Michael}\ \bibnamefont
  {Hermele}}, \ and\ \bibinfo {author} {\bibfnamefont {Rahul}\ \bibnamefont
  {Nandkishore}},\ }\bibfield  {title} {\enquote {\bibinfo {title}
  {Localization from hilbert space shattering: From theory to physical
  realizations},}\ }\href {\doibase 10.1103/PhysRevB.101.174204} {\bibfield
  {journal} {\bibinfo  {journal} {Phys. Rev. B}\ }\textbf {\bibinfo {volume}
  {101}},\ \bibinfo {pages} {174204} (\bibinfo {year} {2020})}\BibitemShut
  {NoStop}%
\bibitem [{\citenamefont {Deutsch}(2018)}]{Deutsch_review}%
  \BibitemOpen
  \bibfield  {author} {\bibinfo {author} {\bibfnamefont {Joshua~M}\
  \bibnamefont {Deutsch}},\ }\bibfield  {title} {\enquote {\bibinfo {title}
  {Eigenstate thermalization hypothesis},}\ }\href {\doibase
  10.1088/1361-6633/aac9f1} {\ \textbf {\bibinfo {volume} {81}},\ \bibinfo
  {pages} {082001} (\bibinfo {year} {2018})}\BibitemShut {NoStop}%
\bibitem [{\citenamefont {D'Alessio}\ \emph {et~al.}(2016)\citenamefont
  {D'Alessio}, \citenamefont {Kafri}, \citenamefont {Polkovnikov},\ and\
  \citenamefont {Rigol}}]{Rigol_review}%
  \BibitemOpen
  \bibfield  {author} {\bibinfo {author} {\bibfnamefont {Luca}\ \bibnamefont
  {D'Alessio}}, \bibinfo {author} {\bibfnamefont {Yariv}\ \bibnamefont
  {Kafri}}, \bibinfo {author} {\bibfnamefont {Anatoli}\ \bibnamefont
  {Polkovnikov}}, \ and\ \bibinfo {author} {\bibfnamefont {Marcos}\
  \bibnamefont {Rigol}},\ }\bibfield  {title} {\enquote {\bibinfo {title} {From
  quantum chaos and eigenstate thermalization to statistical mechanics and
  thermodynamics},}\ }\href {\doibase 10.1080/00018732.2016.1198134} {\bibfield
   {journal} {\bibinfo  {journal} {Advances in Physics}\ }\textbf {\bibinfo
  {volume} {65}},\ \bibinfo {pages} {239--362} (\bibinfo {year} {2016})},\
  \Eprint {http://arxiv.org/abs/https://doi.org/10.1080/00018732.2016.1198134}
  {https://doi.org/10.1080/00018732.2016.1198134} \BibitemShut {NoStop}%
\bibitem [{\citenamefont {Huse}\ \emph {et~al.}(2013)\citenamefont {Huse},
  \citenamefont {Nandkishore}, \citenamefont {Oganesyan}, \citenamefont {Pal},\
  and\ \citenamefont {Sondhi}}]{Huse2013}%
  \BibitemOpen
  \bibfield  {author} {\bibinfo {author} {\bibfnamefont {David~A.}\
  \bibnamefont {Huse}}, \bibinfo {author} {\bibfnamefont {Rahul}\ \bibnamefont
  {Nandkishore}}, \bibinfo {author} {\bibfnamefont {Vadim}\ \bibnamefont
  {Oganesyan}}, \bibinfo {author} {\bibfnamefont {Arijeet}\ \bibnamefont
  {Pal}}, \ and\ \bibinfo {author} {\bibfnamefont {S.~L.}\ \bibnamefont
  {Sondhi}},\ }\bibfield  {title} {\enquote {\bibinfo {title}
  {Localization-protected quantum order},}\ }\href {\doibase
  10.1103/PhysRevB.88.014206} {\bibfield  {journal} {\bibinfo  {journal} {Phys.
  Rev. B}\ }\textbf {\bibinfo {volume} {88}},\ \bibinfo {pages} {014206}
  (\bibinfo {year} {2013})}\BibitemShut {NoStop}%
\bibitem [{\citenamefont {Bauer}\ and\ \citenamefont
  {Nayak}(2013)}]{Bauer2013}%
  \BibitemOpen
  \bibfield  {author} {\bibinfo {author} {\bibfnamefont {Bela}\ \bibnamefont
  {Bauer}}\ and\ \bibinfo {author} {\bibfnamefont {Chetan}\ \bibnamefont
  {Nayak}},\ }\bibfield  {title} {\enquote {\bibinfo {title} {Area laws in a
  many-body localized state and its implications for topological order},}\
  }\href {\doibase 10.1088/1742-5468/2013/09/p09005} {\bibfield  {journal}
  {\bibinfo  {journal} {Journal of Statistical Mechanics: Theory and
  Experiment}\ }\textbf {\bibinfo {volume} {2013}},\ \bibinfo {pages} {P09005}
  (\bibinfo {year} {2013})}\BibitemShut {NoStop}%
\bibitem [{\citenamefont {{Yao}}\ \emph {et~al.}(2015)\citenamefont {{Yao}},
  \citenamefont {{Laumann}},\ and\ \citenamefont {{Vishwanath}}}]{Yao2015}%
  \BibitemOpen
  \bibfield  {author} {\bibinfo {author} {\bibfnamefont {Norman~Y.}\
  \bibnamefont {{Yao}}}, \bibinfo {author} {\bibfnamefont {Chris~R.}\
  \bibnamefont {{Laumann}}}, \ and\ \bibinfo {author} {\bibfnamefont {Ashvin}\
  \bibnamefont {{Vishwanath}}},\ }\bibfield  {title} {\enquote {\bibinfo
  {title} {{Many-body localization protected quantum state transfer}},}\
  }\href@noop {} {\bibfield  {journal} {\bibinfo  {journal} {arXiv e-prints}\
  ,\ \bibinfo {eid} {arXiv:1508.06995}} (\bibinfo {year} {2015})},\ \Eprint
  {http://arxiv.org/abs/1508.06995} {arXiv:1508.06995 [quant-ph]} \BibitemShut
  {NoStop}%
\bibitem [{\citenamefont {von Keyserlingk}\ \emph {et~al.}(2016)\citenamefont
  {von Keyserlingk}, \citenamefont {Khemani},\ and\ \citenamefont
  {Sondhi}}]{vonKeyserlingk2016}%
  \BibitemOpen
  \bibfield  {author} {\bibinfo {author} {\bibfnamefont {C.~W.}\ \bibnamefont
  {von Keyserlingk}}, \bibinfo {author} {\bibfnamefont {Vedika}\ \bibnamefont
  {Khemani}}, \ and\ \bibinfo {author} {\bibfnamefont {S.~L.}\ \bibnamefont
  {Sondhi}},\ }\bibfield  {title} {\enquote {\bibinfo {title} {Absolute
  stability and spatiotemporal long-range order in floquet systems},}\ }\href
  {\doibase 10.1103/PhysRevB.94.085112} {\bibfield  {journal} {\bibinfo
  {journal} {Phys. Rev. B}\ }\textbf {\bibinfo {volume} {94}},\ \bibinfo
  {pages} {085112} (\bibinfo {year} {2016})}\BibitemShut {NoStop}%
\bibitem [{\citenamefont {Khemani}\ \emph {et~al.}(2016)\citenamefont
  {Khemani}, \citenamefont {Lazarides}, \citenamefont {Moessner},\ and\
  \citenamefont {Sondhi}}]{Khemani2016}%
  \BibitemOpen
  \bibfield  {author} {\bibinfo {author} {\bibfnamefont {Vedika}\ \bibnamefont
  {Khemani}}, \bibinfo {author} {\bibfnamefont {Achilleas}\ \bibnamefont
  {Lazarides}}, \bibinfo {author} {\bibfnamefont {Roderich}\ \bibnamefont
  {Moessner}}, \ and\ \bibinfo {author} {\bibfnamefont {S.~L.}\ \bibnamefont
  {Sondhi}},\ }\bibfield  {title} {\enquote {\bibinfo {title} {Phase structure
  of driven quantum systems},}\ }\href {\doibase
  10.1103/PhysRevLett.116.250401} {\bibfield  {journal} {\bibinfo  {journal}
  {Phys. Rev. Lett.}\ }\textbf {\bibinfo {volume} {116}},\ \bibinfo {pages}
  {250401} (\bibinfo {year} {2016})}\BibitemShut {NoStop}%
\bibitem [{\citenamefont {Else}\ \emph {et~al.}(2016)\citenamefont {Else},
  \citenamefont {Bauer},\ and\ \citenamefont {Nayak}}]{Else2016}%
  \BibitemOpen
  \bibfield  {author} {\bibinfo {author} {\bibfnamefont {Dominic~V.}\
  \bibnamefont {Else}}, \bibinfo {author} {\bibfnamefont {Bela}\ \bibnamefont
  {Bauer}}, \ and\ \bibinfo {author} {\bibfnamefont {Chetan}\ \bibnamefont
  {Nayak}},\ }\bibfield  {title} {\enquote {\bibinfo {title} {Floquet time
  crystals},}\ }\href {\doibase 10.1103/PhysRevLett.117.090402} {\bibfield
  {journal} {\bibinfo  {journal} {Phys. Rev. Lett.}\ }\textbf {\bibinfo
  {volume} {117}},\ \bibinfo {pages} {090402} (\bibinfo {year}
  {2016})}\BibitemShut {NoStop}%
\bibitem [{\citenamefont {Yao}\ \emph {et~al.}(2017)\citenamefont {Yao},
  \citenamefont {Potter}, \citenamefont {Potirniche},\ and\ \citenamefont
  {Vishwanath}}]{Yao2017}%
  \BibitemOpen
  \bibfield  {author} {\bibinfo {author} {\bibfnamefont {N.~Y.}\ \bibnamefont
  {Yao}}, \bibinfo {author} {\bibfnamefont {A.~C.}\ \bibnamefont {Potter}},
  \bibinfo {author} {\bibfnamefont {I.-D.}\ \bibnamefont {Potirniche}}, \ and\
  \bibinfo {author} {\bibfnamefont {A.}~\bibnamefont {Vishwanath}},\ }\bibfield
   {title} {\enquote {\bibinfo {title} {Discrete time crystals: Rigidity,
  criticality, and realizations},}\ }\href {\doibase
  10.1103/PhysRevLett.118.030401} {\bibfield  {journal} {\bibinfo  {journal}
  {Phys. Rev. Lett.}\ }\textbf {\bibinfo {volume} {118}},\ \bibinfo {pages}
  {030401} (\bibinfo {year} {2017})}\BibitemShut {NoStop}%
\bibitem [{\citenamefont {Mi}\ \emph {et~al.}(2022)\citenamefont {Mi},
  \citenamefont {Ippoliti}, \citenamefont {Quintana}, \citenamefont {Greene},
  \citenamefont {Chen}, \citenamefont {Gross}, \citenamefont {Arute},
  \citenamefont {Arya}, \citenamefont {Atalaya}, \citenamefont {Babbush},
  \citenamefont {Bardin}, \citenamefont {Basso}, \citenamefont {Bengtsson},
  \citenamefont {Bilmes}, \citenamefont {Bourassa}, \citenamefont {Brill},
  \citenamefont {Broughton}, \citenamefont {Buckley}, \citenamefont {Buell},
  \citenamefont {Burkett}, \citenamefont {Bushnell}, \citenamefont {Chiaro},
  \citenamefont {Collins}, \citenamefont {Courtney}, \citenamefont {Debroy},
  \citenamefont {Demura}, \citenamefont {Derk}, \citenamefont {Dunsworth},
  \citenamefont {Eppens}, \citenamefont {Erickson}, \citenamefont {Farhi},
  \citenamefont {Fowler}, \citenamefont {Foxen}, \citenamefont {Gidney},
  \citenamefont {Giustina}, \citenamefont {Harrigan}, \citenamefont
  {Harrington}, \citenamefont {Hilton}, \citenamefont {Ho}, \citenamefont
  {Hong}, \citenamefont {Huang}, \citenamefont {Huff}, \citenamefont {Huggins},
  \citenamefont {Ioffe}, \citenamefont {Isakov}, \citenamefont {Iveland},
  \citenamefont {Jeffrey}, \citenamefont {Jiang}, \citenamefont {Jones},
  \citenamefont {Kafri}, \citenamefont {Khattar}, \citenamefont {Kim},
  \citenamefont {Kitaev}, \citenamefont {Klimov}, \citenamefont {Korotkov},
  \citenamefont {Kostritsa}, \citenamefont {Landhuis}, \citenamefont {Laptev},
  \citenamefont {Lee}, \citenamefont {Lee}, \citenamefont {Locharla},
  \citenamefont {Lucero}, \citenamefont {Martin}, \citenamefont {McClean},
  \citenamefont {McCourt}, \citenamefont {McEwen}, \citenamefont {Miao},
  \citenamefont {Mohseni}, \citenamefont {Montazeri}, \citenamefont
  {Mruczkiewicz}, \citenamefont {Naaman}, \citenamefont {Neeley}, \citenamefont
  {Neill}, \citenamefont {Newman}, \citenamefont {Niu}, \citenamefont
  {O'Brien}, \citenamefont {Opremcak}, \citenamefont {Ostby}, \citenamefont
  {Pato}, \citenamefont {Petukhov}, \citenamefont {Rubin}, \citenamefont
  {Sank}, \citenamefont {Satzinger}, \citenamefont {Shvarts}, \citenamefont
  {Su}, \citenamefont {Strain}, \citenamefont {Szalay}, \citenamefont
  {Trevithick}, \citenamefont {Villalonga}, \citenamefont {White},
  \citenamefont {Yao}, \citenamefont {Yeh}, \citenamefont {Yoo}, \citenamefont
  {Zalcman}, \citenamefont {Neven}, \citenamefont {Boixo}, \citenamefont
  {Smelyanskiy}, \citenamefont {Megrant}, \citenamefont {Kelly}, \citenamefont
  {Chen}, \citenamefont {Sondhi}, \citenamefont {Moessner}, \citenamefont
  {Kechedzhi}, \citenamefont {Khemani},\ and\ \citenamefont
  {Roushan}}]{Mi2022}%
  \BibitemOpen
  \bibfield  {author} {\bibinfo {author} {\bibfnamefont {Xiao}\ \bibnamefont
  {Mi}}, \bibinfo {author} {\bibfnamefont {Matteo}\ \bibnamefont {Ippoliti}},
  \bibinfo {author} {\bibfnamefont {Chris}\ \bibnamefont {Quintana}}, \bibinfo
  {author} {\bibfnamefont {Ami}\ \bibnamefont {Greene}}, \bibinfo {author}
  {\bibfnamefont {Zijun}\ \bibnamefont {Chen}}, \bibinfo {author}
  {\bibfnamefont {Jonathan}\ \bibnamefont {Gross}}, \bibinfo {author}
  {\bibfnamefont {Frank}\ \bibnamefont {Arute}}, \bibinfo {author}
  {\bibfnamefont {Kunal}\ \bibnamefont {Arya}}, \bibinfo {author}
  {\bibfnamefont {Juan}\ \bibnamefont {Atalaya}}, \bibinfo {author}
  {\bibfnamefont {Ryan}\ \bibnamefont {Babbush}}, \bibinfo {author}
  {\bibfnamefont {Joseph~C.}\ \bibnamefont {Bardin}}, \bibinfo {author}
  {\bibfnamefont {Joao}\ \bibnamefont {Basso}}, \bibinfo {author}
  {\bibfnamefont {Andreas}\ \bibnamefont {Bengtsson}}, \bibinfo {author}
  {\bibfnamefont {Alexander}\ \bibnamefont {Bilmes}}, \bibinfo {author}
  {\bibfnamefont {Alexandre}\ \bibnamefont {Bourassa}}, \bibinfo {author}
  {\bibfnamefont {Leon}\ \bibnamefont {Brill}}, \bibinfo {author}
  {\bibfnamefont {Michael}\ \bibnamefont {Broughton}}, \bibinfo {author}
  {\bibfnamefont {Bob~B.}\ \bibnamefont {Buckley}}, \bibinfo {author}
  {\bibfnamefont {David~A.}\ \bibnamefont {Buell}}, \bibinfo {author}
  {\bibfnamefont {Brian}\ \bibnamefont {Burkett}}, \bibinfo {author}
  {\bibfnamefont {Nicholas}\ \bibnamefont {Bushnell}}, \bibinfo {author}
  {\bibfnamefont {Benjamin}\ \bibnamefont {Chiaro}}, \bibinfo {author}
  {\bibfnamefont {Roberto}\ \bibnamefont {Collins}}, \bibinfo {author}
  {\bibfnamefont {William}\ \bibnamefont {Courtney}}, \bibinfo {author}
  {\bibfnamefont {Dripto}\ \bibnamefont {Debroy}}, \bibinfo {author}
  {\bibfnamefont {Sean}\ \bibnamefont {Demura}}, \bibinfo {author}
  {\bibfnamefont {Alan~R.}\ \bibnamefont {Derk}}, \bibinfo {author}
  {\bibfnamefont {Andrew}\ \bibnamefont {Dunsworth}}, \bibinfo {author}
  {\bibfnamefont {Daniel}\ \bibnamefont {Eppens}}, \bibinfo {author}
  {\bibfnamefont {Catherine}\ \bibnamefont {Erickson}}, \bibinfo {author}
  {\bibfnamefont {Edward}\ \bibnamefont {Farhi}}, \bibinfo {author}
  {\bibfnamefont {Austin~G.}\ \bibnamefont {Fowler}}, \bibinfo {author}
  {\bibfnamefont {Brooks}\ \bibnamefont {Foxen}}, \bibinfo {author}
  {\bibfnamefont {Craig}\ \bibnamefont {Gidney}}, \bibinfo {author}
  {\bibfnamefont {Marissa}\ \bibnamefont {Giustina}}, \bibinfo {author}
  {\bibfnamefont {Matthew~P.}\ \bibnamefont {Harrigan}}, \bibinfo {author}
  {\bibfnamefont {Sean~D.}\ \bibnamefont {Harrington}}, \bibinfo {author}
  {\bibfnamefont {Jeremy}\ \bibnamefont {Hilton}}, \bibinfo {author}
  {\bibfnamefont {Alan}\ \bibnamefont {Ho}}, \bibinfo {author} {\bibfnamefont
  {Sabrina}\ \bibnamefont {Hong}}, \bibinfo {author} {\bibfnamefont {Trent}\
  \bibnamefont {Huang}}, \bibinfo {author} {\bibfnamefont {Ashley}\
  \bibnamefont {Huff}}, \bibinfo {author} {\bibfnamefont {William~J.}\
  \bibnamefont {Huggins}}, \bibinfo {author} {\bibfnamefont {L.~B.}\
  \bibnamefont {Ioffe}}, \bibinfo {author} {\bibfnamefont {Sergei~V.}\
  \bibnamefont {Isakov}}, \bibinfo {author} {\bibfnamefont {Justin}\
  \bibnamefont {Iveland}}, \bibinfo {author} {\bibfnamefont {Evan}\
  \bibnamefont {Jeffrey}}, \bibinfo {author} {\bibfnamefont {Zhang}\
  \bibnamefont {Jiang}}, \bibinfo {author} {\bibfnamefont {Cody}\ \bibnamefont
  {Jones}}, \bibinfo {author} {\bibfnamefont {Dvir}\ \bibnamefont {Kafri}},
  \bibinfo {author} {\bibfnamefont {Tanuj}\ \bibnamefont {Khattar}}, \bibinfo
  {author} {\bibfnamefont {Seon}\ \bibnamefont {Kim}}, \bibinfo {author}
  {\bibfnamefont {Alexei}\ \bibnamefont {Kitaev}}, \bibinfo {author}
  {\bibfnamefont {Paul~V.}\ \bibnamefont {Klimov}}, \bibinfo {author}
  {\bibfnamefont {Alexander~N.}\ \bibnamefont {Korotkov}}, \bibinfo {author}
  {\bibfnamefont {Fedor}\ \bibnamefont {Kostritsa}}, \bibinfo {author}
  {\bibfnamefont {David}\ \bibnamefont {Landhuis}}, \bibinfo {author}
  {\bibfnamefont {Pavel}\ \bibnamefont {Laptev}}, \bibinfo {author}
  {\bibfnamefont {Joonho}\ \bibnamefont {Lee}}, \bibinfo {author}
  {\bibfnamefont {Kenny}\ \bibnamefont {Lee}}, \bibinfo {author} {\bibfnamefont
  {Aditya}\ \bibnamefont {Locharla}}, \bibinfo {author} {\bibfnamefont {Erik}\
  \bibnamefont {Lucero}}, \bibinfo {author} {\bibfnamefont {Orion}\
  \bibnamefont {Martin}}, \bibinfo {author} {\bibfnamefont {Jarrod~R.}\
  \bibnamefont {McClean}}, \bibinfo {author} {\bibfnamefont {Trevor}\
  \bibnamefont {McCourt}}, \bibinfo {author} {\bibfnamefont {Matt}\
  \bibnamefont {McEwen}}, \bibinfo {author} {\bibfnamefont {Kevin~C.}\
  \bibnamefont {Miao}}, \bibinfo {author} {\bibfnamefont {Masoud}\ \bibnamefont
  {Mohseni}}, \bibinfo {author} {\bibfnamefont {Shirin}\ \bibnamefont
  {Montazeri}}, \bibinfo {author} {\bibfnamefont {Wojciech}\ \bibnamefont
  {Mruczkiewicz}}, \bibinfo {author} {\bibfnamefont {Ofer}\ \bibnamefont
  {Naaman}}, \bibinfo {author} {\bibfnamefont {Matthew}\ \bibnamefont
  {Neeley}}, \bibinfo {author} {\bibfnamefont {Charles}\ \bibnamefont {Neill}},
  \bibinfo {author} {\bibfnamefont {Michael}\ \bibnamefont {Newman}}, \bibinfo
  {author} {\bibfnamefont {Murphy~Yuezhen}\ \bibnamefont {Niu}}, \bibinfo
  {author} {\bibfnamefont {Thomas~E.}\ \bibnamefont {O'Brien}}, \bibinfo
  {author} {\bibfnamefont {Alex}\ \bibnamefont {Opremcak}}, \bibinfo {author}
  {\bibfnamefont {Eric}\ \bibnamefont {Ostby}}, \bibinfo {author}
  {\bibfnamefont {Balint}\ \bibnamefont {Pato}}, \bibinfo {author}
  {\bibfnamefont {Andre}\ \bibnamefont {Petukhov}}, \bibinfo {author}
  {\bibfnamefont {Nicholas~C.}\ \bibnamefont {Rubin}}, \bibinfo {author}
  {\bibfnamefont {Daniel}\ \bibnamefont {Sank}}, \bibinfo {author}
  {\bibfnamefont {Kevin~J.}\ \bibnamefont {Satzinger}}, \bibinfo {author}
  {\bibfnamefont {Vladimir}\ \bibnamefont {Shvarts}}, \bibinfo {author}
  {\bibfnamefont {Yuan}\ \bibnamefont {Su}}, \bibinfo {author} {\bibfnamefont
  {Doug}\ \bibnamefont {Strain}}, \bibinfo {author} {\bibfnamefont {Marco}\
  \bibnamefont {Szalay}}, \bibinfo {author} {\bibfnamefont {Matthew~D.}\
  \bibnamefont {Trevithick}}, \bibinfo {author} {\bibfnamefont {Benjamin}\
  \bibnamefont {Villalonga}}, \bibinfo {author} {\bibfnamefont {Theodore}\
  \bibnamefont {White}}, \bibinfo {author} {\bibfnamefont {Z.~Jamie}\
  \bibnamefont {Yao}}, \bibinfo {author} {\bibfnamefont {Ping}\ \bibnamefont
  {Yeh}}, \bibinfo {author} {\bibfnamefont {Juhwan}\ \bibnamefont {Yoo}},
  \bibinfo {author} {\bibfnamefont {Adam}\ \bibnamefont {Zalcman}}, \bibinfo
  {author} {\bibfnamefont {Hartmut}\ \bibnamefont {Neven}}, \bibinfo {author}
  {\bibfnamefont {Sergio}\ \bibnamefont {Boixo}}, \bibinfo {author}
  {\bibfnamefont {Vadim}\ \bibnamefont {Smelyanskiy}}, \bibinfo {author}
  {\bibfnamefont {Anthony}\ \bibnamefont {Megrant}}, \bibinfo {author}
  {\bibfnamefont {Julian}\ \bibnamefont {Kelly}}, \bibinfo {author}
  {\bibfnamefont {Yu}~\bibnamefont {Chen}}, \bibinfo {author} {\bibfnamefont
  {S.~L.}\ \bibnamefont {Sondhi}}, \bibinfo {author} {\bibfnamefont {Roderich}\
  \bibnamefont {Moessner}}, \bibinfo {author} {\bibfnamefont {Kostyantyn}\
  \bibnamefont {Kechedzhi}}, \bibinfo {author} {\bibfnamefont {Vedika}\
  \bibnamefont {Khemani}}, \ and\ \bibinfo {author} {\bibfnamefont {Pedram}\
  \bibnamefont {Roushan}},\ }\bibfield  {title} {\enquote {\bibinfo {title}
  {Time-crystalline eigenstate order on a quantum processor},}\ }\href
  {\doibase 10.1038/s41586-021-04257-w} {\bibfield  {journal} {\bibinfo
  {journal} {Nature}\ }\textbf {\bibinfo {volume} {601}},\ \bibinfo {pages}
  {531--536} (\bibinfo {year} {2022})}\BibitemShut {NoStop}%
\bibitem [{\citenamefont {Smith}\ \emph
  {et~al.}(2017{\natexlab{a}})\citenamefont {Smith}, \citenamefont {Knolle},
  \citenamefont {Kovrizhin},\ and\ \citenamefont {Moessner}}]{Smith2017}%
  \BibitemOpen
  \bibfield  {author} {\bibinfo {author} {\bibfnamefont {A.}~\bibnamefont
  {Smith}}, \bibinfo {author} {\bibfnamefont {J.}~\bibnamefont {Knolle}},
  \bibinfo {author} {\bibfnamefont {D.~L.}\ \bibnamefont {Kovrizhin}}, \ and\
  \bibinfo {author} {\bibfnamefont {R.}~\bibnamefont {Moessner}},\ }\bibfield
  {title} {\enquote {\bibinfo {title} {Disorder-free localization},}\ }\href
  {\doibase 10.1103/PhysRevLett.118.266601} {\bibfield  {journal} {\bibinfo
  {journal} {Phys. Rev. Lett.}\ }\textbf {\bibinfo {volume} {118}},\ \bibinfo
  {pages} {266601} (\bibinfo {year} {2017}{\natexlab{a}})}\BibitemShut
  {NoStop}%
\bibitem [{\citenamefont {Brenes}\ \emph {et~al.}(2018)\citenamefont {Brenes},
  \citenamefont {Dalmonte}, \citenamefont {Heyl},\ and\ \citenamefont
  {Scardicchio}}]{Brenes2018}%
  \BibitemOpen
  \bibfield  {author} {\bibinfo {author} {\bibfnamefont {Marlon}\ \bibnamefont
  {Brenes}}, \bibinfo {author} {\bibfnamefont {Marcello}\ \bibnamefont
  {Dalmonte}}, \bibinfo {author} {\bibfnamefont {Markus}\ \bibnamefont {Heyl}},
  \ and\ \bibinfo {author} {\bibfnamefont {Antonello}\ \bibnamefont
  {Scardicchio}},\ }\bibfield  {title} {\enquote {\bibinfo {title} {Many-body
  localization dynamics from gauge invariance},}\ }\href {\doibase
  10.1103/PhysRevLett.120.030601} {\bibfield  {journal} {\bibinfo  {journal}
  {Phys. Rev. Lett.}\ }\textbf {\bibinfo {volume} {120}},\ \bibinfo {pages}
  {030601} (\bibinfo {year} {2018})}\BibitemShut {NoStop}%
\bibitem [{\citenamefont {Smith}\ \emph
  {et~al.}(2017{\natexlab{b}})\citenamefont {Smith}, \citenamefont {Knolle},
  \citenamefont {Moessner},\ and\ \citenamefont
  {Kovrizhin}}]{smith2017absence}%
  \BibitemOpen
  \bibfield  {author} {\bibinfo {author} {\bibfnamefont {A.}~\bibnamefont
  {Smith}}, \bibinfo {author} {\bibfnamefont {J.}~\bibnamefont {Knolle}},
  \bibinfo {author} {\bibfnamefont {R.}~\bibnamefont {Moessner}}, \ and\
  \bibinfo {author} {\bibfnamefont {D.~L.}\ \bibnamefont {Kovrizhin}},\
  }\bibfield  {title} {\enquote {\bibinfo {title} {Absence of ergodicity
  without quenched disorder: From quantum disentangled liquids to many-body
  localization},}\ }\href {\doibase 10.1103/PhysRevLett.119.176601} {\bibfield
  {journal} {\bibinfo  {journal} {Phys. Rev. Lett.}\ }\textbf {\bibinfo
  {volume} {119}},\ \bibinfo {pages} {176601} (\bibinfo {year}
  {2017}{\natexlab{b}})}\BibitemShut {NoStop}%
\bibitem [{\citenamefont {Metavitsiadis}\ \emph {et~al.}(2017)\citenamefont
  {Metavitsiadis}, \citenamefont {Pidatella},\ and\ \citenamefont
  {Brenig}}]{Metavitsiadis2017}%
  \BibitemOpen
  \bibfield  {author} {\bibinfo {author} {\bibfnamefont {Alexandros}\
  \bibnamefont {Metavitsiadis}}, \bibinfo {author} {\bibfnamefont {Angelo}\
  \bibnamefont {Pidatella}}, \ and\ \bibinfo {author} {\bibfnamefont {Wolfram}\
  \bibnamefont {Brenig}},\ }\bibfield  {title} {\enquote {\bibinfo {title}
  {Thermal transport in a two-dimensional ${\mathbb{z}}_{2}$ spin liquid},}\
  }\href {\doibase 10.1103/PhysRevB.96.205121} {\bibfield  {journal} {\bibinfo
  {journal} {Phys. Rev. B}\ }\textbf {\bibinfo {volume} {96}},\ \bibinfo
  {pages} {205121} (\bibinfo {year} {2017})}\BibitemShut {NoStop}%
\bibitem [{\citenamefont {Smith}\ \emph {et~al.}(2018)\citenamefont {Smith},
  \citenamefont {Knolle}, \citenamefont {Moessner},\ and\ \citenamefont
  {Kovrizhin}}]{Smith2018}%
  \BibitemOpen
  \bibfield  {author} {\bibinfo {author} {\bibfnamefont {Adam}\ \bibnamefont
  {Smith}}, \bibinfo {author} {\bibfnamefont {Johannes}\ \bibnamefont
  {Knolle}}, \bibinfo {author} {\bibfnamefont {Roderich}\ \bibnamefont
  {Moessner}}, \ and\ \bibinfo {author} {\bibfnamefont {Dmitry~L.}\
  \bibnamefont {Kovrizhin}},\ }\bibfield  {title} {\enquote {\bibinfo {title}
  {Dynamical localization in $\mathbb{Z}_2$ lattice gauge theories},}\ }\href
  {\doibase 10.1103/PhysRevB.97.245137} {\bibfield  {journal} {\bibinfo
  {journal} {Phys. Rev. B}\ }\textbf {\bibinfo {volume} {97}},\ \bibinfo
  {pages} {245137} (\bibinfo {year} {2018})}\BibitemShut {NoStop}%
\bibitem [{\citenamefont {Russomanno}\ \emph {et~al.}(2020)\citenamefont
  {Russomanno}, \citenamefont {Notarnicola}, \citenamefont {Surace},
  \citenamefont {Fazio}, \citenamefont {Dalmonte},\ and\ \citenamefont
  {Heyl}}]{Russomanno2020}%
  \BibitemOpen
  \bibfield  {author} {\bibinfo {author} {\bibfnamefont {Angelo}\ \bibnamefont
  {Russomanno}}, \bibinfo {author} {\bibfnamefont {Simone}\ \bibnamefont
  {Notarnicola}}, \bibinfo {author} {\bibfnamefont {Federica~Maria}\
  \bibnamefont {Surace}}, \bibinfo {author} {\bibfnamefont {Rosario}\
  \bibnamefont {Fazio}}, \bibinfo {author} {\bibfnamefont {Marcello}\
  \bibnamefont {Dalmonte}}, \ and\ \bibinfo {author} {\bibfnamefont {Markus}\
  \bibnamefont {Heyl}},\ }\bibfield  {title} {\enquote {\bibinfo {title}
  {Homogeneous floquet time crystal protected by gauge invariance},}\ }\href
  {\doibase 10.1103/PhysRevResearch.2.012003} {\bibfield  {journal} {\bibinfo
  {journal} {Phys. Rev. Research}\ }\textbf {\bibinfo {volume} {2}},\ \bibinfo
  {pages} {012003} (\bibinfo {year} {2020})}\BibitemShut {NoStop}%
\bibitem [{\citenamefont {Papaefstathiou}\ \emph {et~al.}(2020)\citenamefont
  {Papaefstathiou}, \citenamefont {Smith},\ and\ \citenamefont
  {Knolle}}]{Papaefstathiou2020}%
  \BibitemOpen
  \bibfield  {author} {\bibinfo {author} {\bibfnamefont {Irene}\ \bibnamefont
  {Papaefstathiou}}, \bibinfo {author} {\bibfnamefont {Adam}\ \bibnamefont
  {Smith}}, \ and\ \bibinfo {author} {\bibfnamefont {Johannes}\ \bibnamefont
  {Knolle}},\ }\bibfield  {title} {\enquote {\bibinfo {title} {Disorder-free
  localization in a simple $u(1)$ lattice gauge theory},}\ }\href {\doibase
  10.1103/PhysRevB.102.165132} {\bibfield  {journal} {\bibinfo  {journal}
  {Phys. Rev. B}\ }\textbf {\bibinfo {volume} {102}},\ \bibinfo {pages}
  {165132} (\bibinfo {year} {2020})}\BibitemShut {NoStop}%
\bibitem [{\citenamefont {Karpov}\ \emph {et~al.}(2021)\citenamefont {Karpov},
  \citenamefont {Verdel}, \citenamefont {Huang}, \citenamefont {Schmitt},\ and\
  \citenamefont {Heyl}}]{karpov2021disorder}%
  \BibitemOpen
  \bibfield  {author} {\bibinfo {author} {\bibfnamefont {P.}~\bibnamefont
  {Karpov}}, \bibinfo {author} {\bibfnamefont {R.}~\bibnamefont {Verdel}},
  \bibinfo {author} {\bibfnamefont {Y.-P.}\ \bibnamefont {Huang}}, \bibinfo
  {author} {\bibfnamefont {M.}~\bibnamefont {Schmitt}}, \ and\ \bibinfo
  {author} {\bibfnamefont {M.}~\bibnamefont {Heyl}},\ }\bibfield  {title}
  {\enquote {\bibinfo {title} {Disorder-free localization in an interacting 2d
  lattice gauge theory},}\ }\href {\doibase 10.1103/PhysRevLett.126.130401}
  {\bibfield  {journal} {\bibinfo  {journal} {Phys. Rev. Lett.}\ }\textbf
  {\bibinfo {volume} {126}},\ \bibinfo {pages} {130401} (\bibinfo {year}
  {2021})}\BibitemShut {NoStop}%
\bibitem [{\citenamefont {Hart}\ \emph {et~al.}(2021)\citenamefont {Hart},
  \citenamefont {Gopalakrishnan},\ and\ \citenamefont
  {Castelnovo}}]{hart2021logarithmic}%
  \BibitemOpen
  \bibfield  {author} {\bibinfo {author} {\bibfnamefont {Oliver}\ \bibnamefont
  {Hart}}, \bibinfo {author} {\bibfnamefont {Sarang}\ \bibnamefont
  {Gopalakrishnan}}, \ and\ \bibinfo {author} {\bibfnamefont {Claudio}\
  \bibnamefont {Castelnovo}},\ }\bibfield  {title} {\enquote {\bibinfo {title}
  {Logarithmic entanglement growth from disorder-free localization in the
  two-leg compass ladder},}\ }\href {\doibase 10.1103/PhysRevLett.126.227202}
  {\bibfield  {journal} {\bibinfo  {journal} {Phys. Rev. Lett.}\ }\textbf
  {\bibinfo {volume} {126}},\ \bibinfo {pages} {227202} (\bibinfo {year}
  {2021})}\BibitemShut {NoStop}%
\bibitem [{\citenamefont {Zhu}\ and\ \citenamefont {Heyl}(2021)}]{Zhu2021}%
  \BibitemOpen
  \bibfield  {author} {\bibinfo {author} {\bibfnamefont {Guo-Yi}\ \bibnamefont
  {Zhu}}\ and\ \bibinfo {author} {\bibfnamefont {Markus}\ \bibnamefont
  {Heyl}},\ }\bibfield  {title} {\enquote {\bibinfo {title} {Subdiffusive
  dynamics and critical quantum correlations in a disorder-free localized
  kitaev honeycomb model out of equilibrium},}\ }\href {\doibase
  10.1103/PhysRevResearch.3.L032069} {\bibfield  {journal} {\bibinfo  {journal}
  {Phys. Rev. Research}\ }\textbf {\bibinfo {volume} {3}},\ \bibinfo {pages}
  {L032069} (\bibinfo {year} {2021})}\BibitemShut {NoStop}%
\bibitem [{\citenamefont {Sous}\ \emph {et~al.}(2021)\citenamefont {Sous},
  \citenamefont {Kloss}, \citenamefont {Kennes}, \citenamefont {Reichman},\
  and\ \citenamefont {Millis}}]{Sous2021}%
  \BibitemOpen
  \bibfield  {author} {\bibinfo {author} {\bibfnamefont {John}\ \bibnamefont
  {Sous}}, \bibinfo {author} {\bibfnamefont {Benedikt}\ \bibnamefont {Kloss}},
  \bibinfo {author} {\bibfnamefont {Dante~M.}\ \bibnamefont {Kennes}}, \bibinfo
  {author} {\bibfnamefont {David~R.}\ \bibnamefont {Reichman}}, \ and\ \bibinfo
  {author} {\bibfnamefont {Andrew~J.}\ \bibnamefont {Millis}},\ }\bibfield
  {title} {\enquote {\bibinfo {title} {Phonon-induced disorder in dynamics of
  optically pumped metals from nonlinear electron-phonon coupling},}\ }\href
  {\doibase 10.1038/s41467-021-26030-3} {\bibfield  {journal} {\bibinfo
  {journal} {Nature Communications}\ }\textbf {\bibinfo {volume} {12}},\
  \bibinfo {pages} {5803} (\bibinfo {year} {2021})}\BibitemShut {NoStop}%
\bibitem [{\citenamefont {Cirac}\ and\ \citenamefont
  {Zoller}(2012)}]{Cirac2012}%
  \BibitemOpen
  \bibfield  {author} {\bibinfo {author} {\bibfnamefont {J.~Ignacio}\
  \bibnamefont {Cirac}}\ and\ \bibinfo {author} {\bibfnamefont {Peter}\
  \bibnamefont {Zoller}},\ }\bibfield  {title} {\enquote {\bibinfo {title}
  {Goals and opportunities in quantum simulation},}\ }\href {\doibase
  10.1038/nphys2275} {\bibfield  {journal} {\bibinfo  {journal} {Nature
  Physics}\ }\textbf {\bibinfo {volume} {8}},\ \bibinfo {pages} {264--266}
  (\bibinfo {year} {2012})}\BibitemShut {NoStop}%
\bibitem [{\citenamefont {Hauke}\ \emph {et~al.}(2012)\citenamefont {Hauke},
  \citenamefont {Cucchietti}, \citenamefont {Tagliacozzo}, \citenamefont
  {Deutsch},\ and\ \citenamefont {Lewenstein}}]{Hauke2012}%
  \BibitemOpen
  \bibfield  {author} {\bibinfo {author} {\bibfnamefont {Philipp}\ \bibnamefont
  {Hauke}}, \bibinfo {author} {\bibfnamefont {Fernando~M}\ \bibnamefont
  {Cucchietti}}, \bibinfo {author} {\bibfnamefont {Luca}\ \bibnamefont
  {Tagliacozzo}}, \bibinfo {author} {\bibfnamefont {Ivan}\ \bibnamefont
  {Deutsch}}, \ and\ \bibinfo {author} {\bibfnamefont {Maciej}\ \bibnamefont
  {Lewenstein}},\ }\bibfield  {title} {\enquote {\bibinfo {title} {Can one
  trust quantum simulators?}}\ }\href {\doibase 10.1088/0034-4885/75/8/082401}
  {\bibfield  {journal} {\bibinfo  {journal} {Reports on Progress in Physics}\
  }\textbf {\bibinfo {volume} {75}},\ \bibinfo {pages} {082401} (\bibinfo
  {year} {2012})}\BibitemShut {NoStop}%
\bibitem [{\citenamefont {Alexeev}\ \emph {et~al.}(2021)\citenamefont
  {Alexeev}, \citenamefont {Bacon}, \citenamefont {Brown}, \citenamefont
  {Calderbank}, \citenamefont {Carr}, \citenamefont {Chong}, \citenamefont
  {DeMarco}, \citenamefont {Englund}, \citenamefont {Farhi}, \citenamefont
  {Fefferman}, \citenamefont {Gorshkov}, \citenamefont {Houck}, \citenamefont
  {Kim}, \citenamefont {Kimmel}, \citenamefont {Lange}, \citenamefont {Lloyd},
  \citenamefont {Lukin}, \citenamefont {Maslov}, \citenamefont {Maunz},
  \citenamefont {Monroe}, \citenamefont {Preskill}, \citenamefont {Roetteler},
  \citenamefont {Savage},\ and\ \citenamefont {Thompson}}]{Alexeev_review}%
  \BibitemOpen
  \bibfield  {author} {\bibinfo {author} {\bibfnamefont {Yuri}\ \bibnamefont
  {Alexeev}}, \bibinfo {author} {\bibfnamefont {Dave}\ \bibnamefont {Bacon}},
  \bibinfo {author} {\bibfnamefont {Kenneth~R.}\ \bibnamefont {Brown}},
  \bibinfo {author} {\bibfnamefont {Robert}\ \bibnamefont {Calderbank}},
  \bibinfo {author} {\bibfnamefont {Lincoln~D.}\ \bibnamefont {Carr}}, \bibinfo
  {author} {\bibfnamefont {Frederic~T.}\ \bibnamefont {Chong}}, \bibinfo
  {author} {\bibfnamefont {Brian}\ \bibnamefont {DeMarco}}, \bibinfo {author}
  {\bibfnamefont {Dirk}\ \bibnamefont {Englund}}, \bibinfo {author}
  {\bibfnamefont {Edward}\ \bibnamefont {Farhi}}, \bibinfo {author}
  {\bibfnamefont {Bill}\ \bibnamefont {Fefferman}}, \bibinfo {author}
  {\bibfnamefont {Alexey~V.}\ \bibnamefont {Gorshkov}}, \bibinfo {author}
  {\bibfnamefont {Andrew}\ \bibnamefont {Houck}}, \bibinfo {author}
  {\bibfnamefont {Jungsang}\ \bibnamefont {Kim}}, \bibinfo {author}
  {\bibfnamefont {Shelby}\ \bibnamefont {Kimmel}}, \bibinfo {author}
  {\bibfnamefont {Michael}\ \bibnamefont {Lange}}, \bibinfo {author}
  {\bibfnamefont {Seth}\ \bibnamefont {Lloyd}}, \bibinfo {author}
  {\bibfnamefont {Mikhail~D.}\ \bibnamefont {Lukin}}, \bibinfo {author}
  {\bibfnamefont {Dmitri}\ \bibnamefont {Maslov}}, \bibinfo {author}
  {\bibfnamefont {Peter}\ \bibnamefont {Maunz}}, \bibinfo {author}
  {\bibfnamefont {Christopher}\ \bibnamefont {Monroe}}, \bibinfo {author}
  {\bibfnamefont {John}\ \bibnamefont {Preskill}}, \bibinfo {author}
  {\bibfnamefont {Martin}\ \bibnamefont {Roetteler}}, \bibinfo {author}
  {\bibfnamefont {Martin~J.}\ \bibnamefont {Savage}}, \ and\ \bibinfo {author}
  {\bibfnamefont {Jeff}\ \bibnamefont {Thompson}},\ }\href {\doibase
  10.1103/PRXQuantum.2.017001} {\enquote {\bibinfo {title} {Quantum computer
  systems for scientific discovery},}\ } (\bibinfo {year} {2021})\BibitemShut
  {NoStop}%
\bibitem [{\citenamefont {Klco}\ \emph {et~al.}(2022)\citenamefont {Klco},
  \citenamefont {Roggero},\ and\ \citenamefont {Savage}}]{klco2021standard}%
  \BibitemOpen
  \bibfield  {author} {\bibinfo {author} {\bibfnamefont {Natalie}\ \bibnamefont
  {Klco}}, \bibinfo {author} {\bibfnamefont {Alessandro}\ \bibnamefont
  {Roggero}}, \ and\ \bibinfo {author} {\bibfnamefont {Martin~J}\ \bibnamefont
  {Savage}},\ }\bibfield  {title} {\enquote {\bibinfo {title} {Standard model
  physics and the digital quantum revolution: thoughts about the interface},}\
  }\href {\doibase 10.1088/1361-6633/ac58a4} {\bibfield  {journal} {\bibinfo
  {journal} {Reports on Progress in Physics}\ }\textbf {\bibinfo {volume}
  {85}},\ \bibinfo {pages} {064301} (\bibinfo {year} {2022})}\BibitemShut
  {NoStop}%
\bibitem [{\citenamefont {Wu}\ and\ \citenamefont {Hsieh}(2019)}]{Wu2019}%
  \BibitemOpen
  \bibfield  {author} {\bibinfo {author} {\bibfnamefont {Jingxiang}\
  \bibnamefont {Wu}}\ and\ \bibinfo {author} {\bibfnamefont {Timothy~H.}\
  \bibnamefont {Hsieh}},\ }\bibfield  {title} {\enquote {\bibinfo {title}
  {Variational thermal quantum simulation via thermofield double states},}\
  }\href {\doibase 10.1103/PhysRevLett.123.220502} {\bibfield  {journal}
  {\bibinfo  {journal} {Phys. Rev. Lett.}\ }\textbf {\bibinfo {volume} {123}},\
  \bibinfo {pages} {220502} (\bibinfo {year} {2019})}\BibitemShut {NoStop}%
\bibitem [{\citenamefont {Sagastizabal}\ \emph {et~al.}(2021)\citenamefont
  {Sagastizabal}, \citenamefont {Premaratne}, \citenamefont {Klaver},
  \citenamefont {Rol}, \citenamefont {Neg{\^\i}rneac}, \citenamefont {Moreira},
  \citenamefont {Zou}, \citenamefont {Johri}, \citenamefont {Muthusubramanian},
  \citenamefont {Beekman}, \citenamefont {Zachariadis}, \citenamefont
  {Ostroukh}, \citenamefont {Haider}, \citenamefont {Bruno}, \citenamefont
  {Matsuura},\ and\ \citenamefont {DiCarlo}}]{Sagastizabal2021}%
  \BibitemOpen
  \bibfield  {author} {\bibinfo {author} {\bibfnamefont {R.}~\bibnamefont
  {Sagastizabal}}, \bibinfo {author} {\bibfnamefont {S.~P.}\ \bibnamefont
  {Premaratne}}, \bibinfo {author} {\bibfnamefont {B.~A.}\ \bibnamefont
  {Klaver}}, \bibinfo {author} {\bibfnamefont {M.~A.}\ \bibnamefont {Rol}},
  \bibinfo {author} {\bibfnamefont {V.}~\bibnamefont {Neg{\^\i}rneac}},
  \bibinfo {author} {\bibfnamefont {M.~S.}\ \bibnamefont {Moreira}}, \bibinfo
  {author} {\bibfnamefont {X.}~\bibnamefont {Zou}}, \bibinfo {author}
  {\bibfnamefont {S.}~\bibnamefont {Johri}}, \bibinfo {author} {\bibfnamefont
  {N.}~\bibnamefont {Muthusubramanian}}, \bibinfo {author} {\bibfnamefont
  {M.}~\bibnamefont {Beekman}}, \bibinfo {author} {\bibfnamefont
  {C.}~\bibnamefont {Zachariadis}}, \bibinfo {author} {\bibfnamefont {V.~P.}\
  \bibnamefont {Ostroukh}}, \bibinfo {author} {\bibfnamefont {N.}~\bibnamefont
  {Haider}}, \bibinfo {author} {\bibfnamefont {A.}~\bibnamefont {Bruno}},
  \bibinfo {author} {\bibfnamefont {A.~Y.}\ \bibnamefont {Matsuura}}, \ and\
  \bibinfo {author} {\bibfnamefont {L.}~\bibnamefont {DiCarlo}},\ }\bibfield
  {title} {\enquote {\bibinfo {title} {Variational preparation of
  finite-temperature states on a quantum computer},}\ }\href {\doibase
  10.1038/s41534-021-00468-1} {\bibfield  {journal} {\bibinfo  {journal} {npj
  Quantum Information}\ }\textbf {\bibinfo {volume} {7}},\ \bibinfo {pages}
  {130} (\bibinfo {year} {2021})}\BibitemShut {NoStop}%
\bibitem [{\citenamefont {Zhang}\ \emph {et~al.}(2021)\citenamefont {Zhang},
  \citenamefont {Zhang}, \citenamefont {Xue}, \citenamefont {Zhu},\ and\
  \citenamefont {Wang}}]{Zhang2021CV}%
  \BibitemOpen
  \bibfield  {author} {\bibinfo {author} {\bibfnamefont {Dan-Bo}\ \bibnamefont
  {Zhang}}, \bibinfo {author} {\bibfnamefont {Guo-Qing}\ \bibnamefont {Zhang}},
  \bibinfo {author} {\bibfnamefont {Zheng-Yuan}\ \bibnamefont {Xue}}, \bibinfo
  {author} {\bibfnamefont {Shi-Liang}\ \bibnamefont {Zhu}}, \ and\ \bibinfo
  {author} {\bibfnamefont {Z.~D.}\ \bibnamefont {Wang}},\ }\bibfield  {title}
  {\enquote {\bibinfo {title} {Continuous-variable assisted thermal quantum
  simulation},}\ }\href {\doibase 10.1103/PhysRevLett.127.020502} {\bibfield
  {journal} {\bibinfo  {journal} {Phys. Rev. Lett.}\ }\textbf {\bibinfo
  {volume} {127}},\ \bibinfo {pages} {020502} (\bibinfo {year}
  {2021})}\BibitemShut {NoStop}%
\bibitem [{\citenamefont {Lu}\ \emph {et~al.}(2021)\citenamefont {Lu},
  \citenamefont {Ba\~nuls},\ and\ \citenamefont {Cirac}}]{Lu2022}%
  \BibitemOpen
  \bibfield  {author} {\bibinfo {author} {\bibfnamefont {Sirui}\ \bibnamefont
  {Lu}}, \bibinfo {author} {\bibfnamefont {Mari~Carmen}\ \bibnamefont
  {Ba\~nuls}}, \ and\ \bibinfo {author} {\bibfnamefont {J.~Ignacio}\
  \bibnamefont {Cirac}},\ }\bibfield  {title} {\enquote {\bibinfo {title}
  {Algorithms for quantum simulation at finite energies},}\ }\href {\doibase
  10.1103/PRXQuantum.2.020321} {\bibfield  {journal} {\bibinfo  {journal} {PRX
  Quantum}\ }\textbf {\bibinfo {volume} {2}},\ \bibinfo {pages} {020321}
  (\bibinfo {year} {2021})}\BibitemShut {NoStop}%
\bibitem [{\citenamefont {{Schuckert}}\ \emph {et~al.}(2022)\citenamefont
  {{Schuckert}}, \citenamefont {{Bohrdt}}, \citenamefont {{Crane}},\ and\
  \citenamefont {{Knap}}}]{Schuckert2022}%
  \BibitemOpen
  \bibfield  {author} {\bibinfo {author} {\bibfnamefont {Alexander}\
  \bibnamefont {{Schuckert}}}, \bibinfo {author} {\bibfnamefont {Annabelle}\
  \bibnamefont {{Bohrdt}}}, \bibinfo {author} {\bibfnamefont {Eleanor}\
  \bibnamefont {{Crane}}}, \ and\ \bibinfo {author} {\bibfnamefont {Michael}\
  \bibnamefont {{Knap}}},\ }\bibfield  {title} {\enquote {\bibinfo {title}
  {{Probing finite-temperature observables in quantum simulators with
  short-time dynamics}},}\ }\href@noop {} {\bibfield  {journal} {\bibinfo
  {journal} {arXiv e-prints}\ ,\ \bibinfo {eid} {arXiv:2206.01756}} (\bibinfo
  {year} {2022})},\ \Eprint {http://arxiv.org/abs/2206.01756} {arXiv:2206.01756
  [quant-ph]} \BibitemShut {NoStop}%
\bibitem [{\citenamefont {Halimeh}\ \emph
  {et~al.}(2021{\natexlab{a}})\citenamefont {Halimeh}, \citenamefont {Homeier},
  \citenamefont {Schweizer}, \citenamefont {Aidelsburger}, \citenamefont
  {Hauke},\ and\ \citenamefont {Grusdt}}]{Halimeh2021stabilizing}%
  \BibitemOpen
  \bibfield  {author} {\bibinfo {author} {\bibfnamefont {Jad~C.}\ \bibnamefont
  {Halimeh}}, \bibinfo {author} {\bibfnamefont {Lukas}\ \bibnamefont
  {Homeier}}, \bibinfo {author} {\bibfnamefont {Christian}\ \bibnamefont
  {Schweizer}}, \bibinfo {author} {\bibfnamefont {Monika}\ \bibnamefont
  {Aidelsburger}}, \bibinfo {author} {\bibfnamefont {Philipp}\ \bibnamefont
  {Hauke}}, \ and\ \bibinfo {author} {\bibfnamefont {Fabian}\ \bibnamefont
  {Grusdt}},\ }\bibfield  {title} {\enquote {\bibinfo {title} {Stabilizing
  lattice gauge theories through simplified local pseudo generators},}\
  }\href@noop {} {\  (\bibinfo {year} {2021}{\natexlab{a}})},\ \Eprint
  {http://arxiv.org/abs/2108.02203} {arXiv:2108.02203 [cond-mat.quant-gas]}
  \BibitemShut {NoStop}%
\bibitem [{\citenamefont {Halimeh}\ \emph {et~al.}(2022)\citenamefont
  {Halimeh}, \citenamefont {Homeier}, \citenamefont {Zhao}, \citenamefont
  {Bohrdt}, \citenamefont {Grusdt}, \citenamefont {Hauke},\ and\ \citenamefont
  {Knolle}}]{Halimeh2021enhancing}%
  \BibitemOpen
  \bibfield  {author} {\bibinfo {author} {\bibfnamefont {Jad~C.}\ \bibnamefont
  {Halimeh}}, \bibinfo {author} {\bibfnamefont {Lukas}\ \bibnamefont
  {Homeier}}, \bibinfo {author} {\bibfnamefont {Hongzheng}\ \bibnamefont
  {Zhao}}, \bibinfo {author} {\bibfnamefont {Annabelle}\ \bibnamefont
  {Bohrdt}}, \bibinfo {author} {\bibfnamefont {Fabian}\ \bibnamefont {Grusdt}},
  \bibinfo {author} {\bibfnamefont {Philipp}\ \bibnamefont {Hauke}}, \ and\
  \bibinfo {author} {\bibfnamefont {Johannes}\ \bibnamefont {Knolle}},\
  }\bibfield  {title} {\enquote {\bibinfo {title} {Enhancing disorder-free
  localization through dynamically emergent local symmetries},}\ }\href
  {\doibase 10.1103/PRXQuantum.3.020345} {\bibfield  {journal} {\bibinfo
  {journal} {PRX Quantum}\ }\textbf {\bibinfo {volume} {3}},\ \bibinfo {pages}
  {020345} (\bibinfo {year} {2022})}\BibitemShut {NoStop}%
\bibitem [{\citenamefont {Rothe}(2005)}]{Rothe_book}%
  \BibitemOpen
  \bibfield  {author} {\bibinfo {author} {\bibfnamefont {H.J.}\ \bibnamefont
  {Rothe}},\ }\href {https://books.google.de/books?id=U1hBLG-\_WxAC} {\emph
  {\bibinfo {title} {Lattice Gauge Theories: An Introduction}}},\ EBSCO ebook
  academic collection\ (\bibinfo  {publisher} {World Scientific},\ \bibinfo
  {year} {2005})\BibitemShut {NoStop}%
\bibitem [{\citenamefont {Zee}(2003)}]{Zee_book}%
  \BibitemOpen
  \bibfield  {author} {\bibinfo {author} {\bibfnamefont {A.}~\bibnamefont
  {Zee}},\ }\href {https://books.google.de/books?id=85G9QgAACAAJ} {\emph
  {\bibinfo {title} {Quantum Field Theory in a Nutshell}}}\ (\bibinfo
  {publisher} {Princeton University Press},\ \bibinfo {year}
  {2003})\BibitemShut {NoStop}%
\bibitem [{\citenamefont {Chandrasekharan}\ and\ \citenamefont
  {Wiese}(1997)}]{Chandrasekharan1997}%
  \BibitemOpen
  \bibfield  {author} {\bibinfo {author} {\bibfnamefont {S}~\bibnamefont
  {Chandrasekharan}}\ and\ \bibinfo {author} {\bibfnamefont {U.-J}\
  \bibnamefont {Wiese}},\ }\bibfield  {title} {\enquote {\bibinfo {title}
  {Quantum link models: A discrete approach to gauge theories},}\ }\href
  {\doibase https://doi.org/10.1016/S0550-3213(97)80041-7} {\bibfield
  {journal} {\bibinfo  {journal} {Nuclear Physics B}\ }\textbf {\bibinfo
  {volume} {492}},\ \bibinfo {pages} {455 -- 471} (\bibinfo {year}
  {1997})}\BibitemShut {NoStop}%
\bibitem [{\citenamefont {Wiese}(2013)}]{Wiese_review}%
  \BibitemOpen
  \bibfield  {author} {\bibinfo {author} {\bibfnamefont {U.-J.}\ \bibnamefont
  {Wiese}},\ }\bibfield  {title} {\enquote {\bibinfo {title} {Ultracold quantum
  gases and lattice systems: quantum simulation of lattice gauge theories},}\
  }\href {\doibase 10.1002/andp.201300104} {\bibfield  {journal} {\bibinfo
  {journal} {Annalen der Physik}\ }\textbf {\bibinfo {volume} {525}},\ \bibinfo
  {pages} {777--796} (\bibinfo {year} {2013})}\BibitemShut {NoStop}%
\bibitem [{\citenamefont {Hauke}\ \emph {et~al.}(2013)\citenamefont {Hauke},
  \citenamefont {Marcos}, \citenamefont {Dalmonte},\ and\ \citenamefont
  {Zoller}}]{Hauke2013}%
  \BibitemOpen
  \bibfield  {author} {\bibinfo {author} {\bibfnamefont {P.}~\bibnamefont
  {Hauke}}, \bibinfo {author} {\bibfnamefont {D.}~\bibnamefont {Marcos}},
  \bibinfo {author} {\bibfnamefont {M.}~\bibnamefont {Dalmonte}}, \ and\
  \bibinfo {author} {\bibfnamefont {P.}~\bibnamefont {Zoller}},\ }\bibfield
  {title} {\enquote {\bibinfo {title} {Quantum simulation of a lattice
  schwinger model in a chain of trapped ions},}\ }\href {\doibase
  10.1103/PhysRevX.3.041018} {\bibfield  {journal} {\bibinfo  {journal} {Phys.
  Rev. X}\ }\textbf {\bibinfo {volume} {3}},\ \bibinfo {pages} {041018}
  (\bibinfo {year} {2013})}\BibitemShut {NoStop}%
\bibitem [{\citenamefont {Yang}\ \emph {et~al.}(2016)\citenamefont {Yang},
  \citenamefont {Giri}, \citenamefont {Johanning}, \citenamefont {Wunderlich},
  \citenamefont {Zoller},\ and\ \citenamefont {Hauke}}]{Yang2016}%
  \BibitemOpen
  \bibfield  {author} {\bibinfo {author} {\bibfnamefont {Dayou}\ \bibnamefont
  {Yang}}, \bibinfo {author} {\bibfnamefont {Gouri~Shankar}\ \bibnamefont
  {Giri}}, \bibinfo {author} {\bibfnamefont {Michael}\ \bibnamefont
  {Johanning}}, \bibinfo {author} {\bibfnamefont {Christof}\ \bibnamefont
  {Wunderlich}}, \bibinfo {author} {\bibfnamefont {Peter}\ \bibnamefont
  {Zoller}}, \ and\ \bibinfo {author} {\bibfnamefont {Philipp}\ \bibnamefont
  {Hauke}},\ }\bibfield  {title} {\enquote {\bibinfo {title} {Analog quantum
  simulation of $(1+1)$-dimensional lattice qed with trapped ions},}\ }\href
  {\doibase 10.1103/PhysRevA.94.052321} {\bibfield  {journal} {\bibinfo
  {journal} {Phys. Rev. A}\ }\textbf {\bibinfo {volume} {94}},\ \bibinfo
  {pages} {052321} (\bibinfo {year} {2016})}\BibitemShut {NoStop}%
\bibitem [{\citenamefont {Kasper}\ \emph {et~al.}(2017)\citenamefont {Kasper},
  \citenamefont {Hebenstreit}, \citenamefont {Jendrzejewski}, \citenamefont
  {Oberthaler},\ and\ \citenamefont {Berges}}]{Kasper2017}%
  \BibitemOpen
  \bibfield  {author} {\bibinfo {author} {\bibfnamefont {V}~\bibnamefont
  {Kasper}}, \bibinfo {author} {\bibfnamefont {F}~\bibnamefont {Hebenstreit}},
  \bibinfo {author} {\bibfnamefont {F}~\bibnamefont {Jendrzejewski}}, \bibinfo
  {author} {\bibfnamefont {M~K}\ \bibnamefont {Oberthaler}}, \ and\ \bibinfo
  {author} {\bibfnamefont {J}~\bibnamefont {Berges}},\ }\bibfield  {title}
  {\enquote {\bibinfo {title} {Implementing quantum electrodynamics with
  ultracold atomic systems},}\ }\href {\doibase 10.1088/1367-2630/aa54e0}
  {\bibfield  {journal} {\bibinfo  {journal} {New Journal of Physics}\ }\textbf
  {\bibinfo {volume} {19}},\ \bibinfo {pages} {023030} (\bibinfo {year}
  {2017})}\BibitemShut {NoStop}%
\bibitem [{\citenamefont {Zohar}\ \emph {et~al.}(2017)\citenamefont {Zohar},
  \citenamefont {Farace}, \citenamefont {Reznik},\ and\ \citenamefont
  {Cirac}}]{Zohar2017}%
  \BibitemOpen
  \bibfield  {author} {\bibinfo {author} {\bibfnamefont {Erez}\ \bibnamefont
  {Zohar}}, \bibinfo {author} {\bibfnamefont {Alessandro}\ \bibnamefont
  {Farace}}, \bibinfo {author} {\bibfnamefont {Benni}\ \bibnamefont {Reznik}},
  \ and\ \bibinfo {author} {\bibfnamefont {J.~Ignacio}\ \bibnamefont {Cirac}},\
  }\bibfield  {title} {\enquote {\bibinfo {title} {Digital quantum simulation
  of ${\mathbb{z}}_{2}$ lattice gauge theories with dynamical fermionic
  matter},}\ }\href {\doibase 10.1103/PhysRevLett.118.070501} {\bibfield
  {journal} {\bibinfo  {journal} {Phys. Rev. Lett.}\ }\textbf {\bibinfo
  {volume} {118}},\ \bibinfo {pages} {070501} (\bibinfo {year}
  {2017})}\BibitemShut {NoStop}%
\bibitem [{\citenamefont {Borla}\ \emph {et~al.}(2020)\citenamefont {Borla},
  \citenamefont {Verresen}, \citenamefont {Grusdt},\ and\ \citenamefont
  {Moroz}}]{Borla2019}%
  \BibitemOpen
  \bibfield  {author} {\bibinfo {author} {\bibfnamefont {Umberto}\ \bibnamefont
  {Borla}}, \bibinfo {author} {\bibfnamefont {Ruben}\ \bibnamefont {Verresen}},
  \bibinfo {author} {\bibfnamefont {Fabian}\ \bibnamefont {Grusdt}}, \ and\
  \bibinfo {author} {\bibfnamefont {Sergej}\ \bibnamefont {Moroz}},\ }\bibfield
   {title} {\enquote {\bibinfo {title} {Confined phases of one-dimensional
  spinless fermions coupled to ${Z}_{2}$ gauge theory},}\ }\href {\doibase
  10.1103/PhysRevLett.124.120503} {\bibfield  {journal} {\bibinfo  {journal}
  {Phys. Rev. Lett.}\ }\textbf {\bibinfo {volume} {124}},\ \bibinfo {pages}
  {120503} (\bibinfo {year} {2020})}\BibitemShut {NoStop}%
\bibitem [{\citenamefont {Yang}\ \emph
  {et~al.}(2020{\natexlab{a}})\citenamefont {Yang}, \citenamefont {Liu},
  \citenamefont {Gorshkov},\ and\ \citenamefont
  {Iadecola}}]{Yang2020fragmentation}%
  \BibitemOpen
  \bibfield  {author} {\bibinfo {author} {\bibfnamefont {Zhi-Cheng}\
  \bibnamefont {Yang}}, \bibinfo {author} {\bibfnamefont {Fangli}\ \bibnamefont
  {Liu}}, \bibinfo {author} {\bibfnamefont {Alexey~V.}\ \bibnamefont
  {Gorshkov}}, \ and\ \bibinfo {author} {\bibfnamefont {Thomas}\ \bibnamefont
  {Iadecola}},\ }\bibfield  {title} {\enquote {\bibinfo {title} {Hilbert-space
  fragmentation from strict confinement},}\ }\href {\doibase
  10.1103/PhysRevLett.124.207602} {\bibfield  {journal} {\bibinfo  {journal}
  {Phys. Rev. Lett.}\ }\textbf {\bibinfo {volume} {124}},\ \bibinfo {pages}
  {207602} (\bibinfo {year} {2020}{\natexlab{a}})}\BibitemShut {NoStop}%
\bibitem [{\citenamefont {Kebri\ifmmode~\check{c}\else \v{c}\fi{}}\ \emph
  {et~al.}(2021)\citenamefont {Kebri\ifmmode~\check{c}\else \v{c}\fi{}},
  \citenamefont {Barbiero}, \citenamefont {Reinmoser}, \citenamefont
  {Schollw\"ock},\ and\ \citenamefont {Grusdt}}]{kebric2021confinement}%
  \BibitemOpen
  \bibfield  {author} {\bibinfo {author} {\bibfnamefont {Matja\ifmmode
  \check{z}\else~\v{z}\fi{}}\ \bibnamefont {Kebri\ifmmode~\check{c}\else
  \v{c}\fi{}}}, \bibinfo {author} {\bibfnamefont {Luca}\ \bibnamefont
  {Barbiero}}, \bibinfo {author} {\bibfnamefont {Christian}\ \bibnamefont
  {Reinmoser}}, \bibinfo {author} {\bibfnamefont {Ulrich}\ \bibnamefont
  {Schollw\"ock}}, \ and\ \bibinfo {author} {\bibfnamefont {Fabian}\
  \bibnamefont {Grusdt}},\ }\bibfield  {title} {\enquote {\bibinfo {title}
  {Confinement and mott transitions of dynamical charges in one-dimensional
  lattice gauge theories},}\ }\href {\doibase 10.1103/PhysRevLett.127.167203}
  {\bibfield  {journal} {\bibinfo  {journal} {Phys. Rev. Lett.}\ }\textbf
  {\bibinfo {volume} {127}},\ \bibinfo {pages} {167203} (\bibinfo {year}
  {2021})}\BibitemShut {NoStop}%
\bibitem [{\citenamefont {Martinez}\ \emph {et~al.}(2016)\citenamefont
  {Martinez}, \citenamefont {Muschik}, \citenamefont {Schindler}, \citenamefont
  {Nigg}, \citenamefont {Erhard}, \citenamefont {Heyl}, \citenamefont {Hauke},
  \citenamefont {Dalmonte}, \citenamefont {Monz}, \citenamefont {Zoller},\ and\
  \citenamefont {Blatt}}]{Martinez2016}%
  \BibitemOpen
  \bibfield  {author} {\bibinfo {author} {\bibfnamefont {Esteban~A.}\
  \bibnamefont {Martinez}}, \bibinfo {author} {\bibfnamefont {Christine~A.}\
  \bibnamefont {Muschik}}, \bibinfo {author} {\bibfnamefont {Philipp}\
  \bibnamefont {Schindler}}, \bibinfo {author} {\bibfnamefont {Daniel}\
  \bibnamefont {Nigg}}, \bibinfo {author} {\bibfnamefont {Alexander}\
  \bibnamefont {Erhard}}, \bibinfo {author} {\bibfnamefont {Markus}\
  \bibnamefont {Heyl}}, \bibinfo {author} {\bibfnamefont {Philipp}\
  \bibnamefont {Hauke}}, \bibinfo {author} {\bibfnamefont {Marcello}\
  \bibnamefont {Dalmonte}}, \bibinfo {author} {\bibfnamefont {Thomas}\
  \bibnamefont {Monz}}, \bibinfo {author} {\bibfnamefont {Peter}\ \bibnamefont
  {Zoller}}, \ and\ \bibinfo {author} {\bibfnamefont {Rainer}\ \bibnamefont
  {Blatt}},\ }\bibfield  {title} {\enquote {\bibinfo {title} {Real-time
  dynamics of lattice gauge theories with a few-qubit quantum computer},}\
  }\href {\doibase 10.1038/nature18318} {\bibfield  {journal} {\bibinfo
  {journal} {Nature}\ }\textbf {\bibinfo {volume} {534}},\ \bibinfo {pages}
  {516--519} (\bibinfo {year} {2016})}\BibitemShut {NoStop}%
\bibitem [{\citenamefont {G{\"o}rg}\ \emph {et~al.}(2019)\citenamefont
  {G{\"o}rg}, \citenamefont {Sandholzer}, \citenamefont {Minguzzi},
  \citenamefont {Desbuquois}, \citenamefont {Messer},\ and\ \citenamefont
  {Esslinger}}]{Goerg2019}%
  \BibitemOpen
  \bibfield  {author} {\bibinfo {author} {\bibfnamefont {Frederik}\
  \bibnamefont {G{\"o}rg}}, \bibinfo {author} {\bibfnamefont {Kilian}\
  \bibnamefont {Sandholzer}}, \bibinfo {author} {\bibfnamefont {Joaqu{\'\i}n}\
  \bibnamefont {Minguzzi}}, \bibinfo {author} {\bibfnamefont {R{\'e}mi}\
  \bibnamefont {Desbuquois}}, \bibinfo {author} {\bibfnamefont {Michael}\
  \bibnamefont {Messer}}, \ and\ \bibinfo {author} {\bibfnamefont {Tilman}\
  \bibnamefont {Esslinger}},\ }\bibfield  {title} {\enquote {\bibinfo {title}
  {Realization of density-dependent peierls phases to engineer quantized gauge
  fields coupled to ultracold matter},}\ }\href {\doibase
  10.1038/s41567-019-0615-4} {\bibfield  {journal} {\bibinfo  {journal} {Nature
  Physics}\ }\textbf {\bibinfo {volume} {15}},\ \bibinfo {pages} {1161--1167}
  (\bibinfo {year} {2019})}\BibitemShut {NoStop}%
\bibitem [{\citenamefont {Schweizer}\ \emph {et~al.}(2019)\citenamefont
  {Schweizer}, \citenamefont {Grusdt}, \citenamefont {Berngruber},
  \citenamefont {Barbiero}, \citenamefont {Demler}, \citenamefont {Goldman},
  \citenamefont {Bloch},\ and\ \citenamefont {Aidelsburger}}]{Schweizer2019}%
  \BibitemOpen
  \bibfield  {author} {\bibinfo {author} {\bibfnamefont {Christian}\
  \bibnamefont {Schweizer}}, \bibinfo {author} {\bibfnamefont {Fabian}\
  \bibnamefont {Grusdt}}, \bibinfo {author} {\bibfnamefont {Moritz}\
  \bibnamefont {Berngruber}}, \bibinfo {author} {\bibfnamefont {Luca}\
  \bibnamefont {Barbiero}}, \bibinfo {author} {\bibfnamefont {Eugene}\
  \bibnamefont {Demler}}, \bibinfo {author} {\bibfnamefont {Nathan}\
  \bibnamefont {Goldman}}, \bibinfo {author} {\bibfnamefont {Immanuel}\
  \bibnamefont {Bloch}}, \ and\ \bibinfo {author} {\bibfnamefont {Monika}\
  \bibnamefont {Aidelsburger}},\ }\bibfield  {title} {\enquote {\bibinfo
  {title} {Floquet approach to $\mathbb{Z}$2 lattice gauge theories with
  ultracold atoms in optical lattices},}\ }\href {\doibase
  10.1038/s41567-019-0649-7} {\bibfield  {journal} {\bibinfo  {journal} {Nature
  Physics}\ }\textbf {\bibinfo {volume} {15}},\ \bibinfo {pages} {1168--1173}
  (\bibinfo {year} {2019})}\BibitemShut {NoStop}%
\bibitem [{\citenamefont {Yang}\ \emph
  {et~al.}(2020{\natexlab{b}})\citenamefont {Yang}, \citenamefont {Sun},
  \citenamefont {Ott}, \citenamefont {Wang}, \citenamefont {Zache},
  \citenamefont {Halimeh}, \citenamefont {Yuan}, \citenamefont {Hauke},\ and\
  \citenamefont {Pan}}]{Yang2020}%
  \BibitemOpen
  \bibfield  {author} {\bibinfo {author} {\bibfnamefont {Bing}\ \bibnamefont
  {Yang}}, \bibinfo {author} {\bibfnamefont {Hui}\ \bibnamefont {Sun}},
  \bibinfo {author} {\bibfnamefont {Robert}\ \bibnamefont {Ott}}, \bibinfo
  {author} {\bibfnamefont {Han-Yi}\ \bibnamefont {Wang}}, \bibinfo {author}
  {\bibfnamefont {Torsten~V.}\ \bibnamefont {Zache}}, \bibinfo {author}
  {\bibfnamefont {Jad~C.}\ \bibnamefont {Halimeh}}, \bibinfo {author}
  {\bibfnamefont {Zhen-Sheng}\ \bibnamefont {Yuan}}, \bibinfo {author}
  {\bibfnamefont {Philipp}\ \bibnamefont {Hauke}}, \ and\ \bibinfo {author}
  {\bibfnamefont {Jian-Wei}\ \bibnamefont {Pan}},\ }\bibfield  {title}
  {\enquote {\bibinfo {title} {Observation of gauge invariance in a 71-site
  bose--hubbard quantum simulator},}\ }\href {\doibase
  10.1038/s41586-020-2910-8} {\bibfield  {journal} {\bibinfo  {journal}
  {Nature}\ }\textbf {\bibinfo {volume} {587}},\ \bibinfo {pages} {392--396}
  (\bibinfo {year} {2020}{\natexlab{b}})}\BibitemShut {NoStop}%
\bibitem [{\citenamefont {Zhou}\ \emph {et~al.}(2021)\citenamefont {Zhou},
  \citenamefont {Su}, \citenamefont {Halimeh}, \citenamefont {Ott},
  \citenamefont {Sun}, \citenamefont {Hauke}, \citenamefont {Yang},
  \citenamefont {Yuan}, \citenamefont {Berges},\ and\ \citenamefont
  {Pan}}]{Zhou2021}%
  \BibitemOpen
  \bibfield  {author} {\bibinfo {author} {\bibfnamefont {Zhao-Yu}\ \bibnamefont
  {Zhou}}, \bibinfo {author} {\bibfnamefont {Guo-Xian}\ \bibnamefont {Su}},
  \bibinfo {author} {\bibfnamefont {Jad~C.}\ \bibnamefont {Halimeh}}, \bibinfo
  {author} {\bibfnamefont {Robert}\ \bibnamefont {Ott}}, \bibinfo {author}
  {\bibfnamefont {Hui}\ \bibnamefont {Sun}}, \bibinfo {author} {\bibfnamefont
  {Philipp}\ \bibnamefont {Hauke}}, \bibinfo {author} {\bibfnamefont {Bing}\
  \bibnamefont {Yang}}, \bibinfo {author} {\bibfnamefont {Zhen-Sheng}\
  \bibnamefont {Yuan}}, \bibinfo {author} {\bibfnamefont {Jürgen}\
  \bibnamefont {Berges}}, \ and\ \bibinfo {author} {\bibfnamefont {Jian-Wei}\
  \bibnamefont {Pan}},\ }\bibfield  {title} {\enquote {\bibinfo {title}
  {Thermalization dynamics of a gauge theory on a quantum simulator},}\
  }\href@noop {} {\  (\bibinfo {year} {2021})},\ \Eprint
  {http://arxiv.org/abs/2107.13563} {arXiv:2107.13563 [cond-mat.quant-gas]}
  \BibitemShut {NoStop}%
\bibitem [{\citenamefont {Wang}\ \emph {et~al.}(2022)\citenamefont {Wang},
  \citenamefont {Ge}, \citenamefont {Xiang}, \citenamefont {Song},
  \citenamefont {Huang}, \citenamefont {Song}, \citenamefont {Guo},
  \citenamefont {Su}, \citenamefont {Xu}, \citenamefont {Zheng},\ and\
  \citenamefont {Fan}}]{Wang2022}%
  \BibitemOpen
  \bibfield  {author} {\bibinfo {author} {\bibfnamefont {Zhan}\ \bibnamefont
  {Wang}}, \bibinfo {author} {\bibfnamefont {Zi-Yong}\ \bibnamefont {Ge}},
  \bibinfo {author} {\bibfnamefont {Zhongcheng}\ \bibnamefont {Xiang}},
  \bibinfo {author} {\bibfnamefont {Xiaohui}\ \bibnamefont {Song}}, \bibinfo
  {author} {\bibfnamefont {Rui-Zhen}\ \bibnamefont {Huang}}, \bibinfo {author}
  {\bibfnamefont {Pengtao}\ \bibnamefont {Song}}, \bibinfo {author}
  {\bibfnamefont {Xue-Yi}\ \bibnamefont {Guo}}, \bibinfo {author}
  {\bibfnamefont {Luhong}\ \bibnamefont {Su}}, \bibinfo {author} {\bibfnamefont
  {Kai}\ \bibnamefont {Xu}}, \bibinfo {author} {\bibfnamefont {Dongning}\
  \bibnamefont {Zheng}}, \ and\ \bibinfo {author} {\bibfnamefont {Heng}\
  \bibnamefont {Fan}},\ }\bibfield  {title} {\enquote {\bibinfo {title}
  {Observation of emergent $\mathbb{Z}_2$ gauge invariance in a superconducting
  circuit},}\ }\href {\doibase 10.1103/PhysRevResearch.4.L022060} {\bibfield
  {journal} {\bibinfo  {journal} {Phys. Rev. Research}\ }\textbf {\bibinfo
  {volume} {4}},\ \bibinfo {pages} {L022060} (\bibinfo {year}
  {2022})}\BibitemShut {NoStop}%
\bibitem [{\citenamefont {{Mildenberger}}\ \emph {et~al.}(2022)\citenamefont
  {{Mildenberger}}, \citenamefont {{Mruczkiewicz}}, \citenamefont {{Halimeh}},
  \citenamefont {{Jiang}},\ and\ \citenamefont {{Hauke}}}]{Mildenberger2022}%
  \BibitemOpen
  \bibfield  {author} {\bibinfo {author} {\bibfnamefont {Julius}\ \bibnamefont
  {{Mildenberger}}}, \bibinfo {author} {\bibfnamefont {Wojciech}\ \bibnamefont
  {{Mruczkiewicz}}}, \bibinfo {author} {\bibfnamefont {Jad~C.}\ \bibnamefont
  {{Halimeh}}}, \bibinfo {author} {\bibfnamefont {Zhang}\ \bibnamefont
  {{Jiang}}}, \ and\ \bibinfo {author} {\bibfnamefont {Philipp}\ \bibnamefont
  {{Hauke}}},\ }\bibfield  {title} {\enquote {\bibinfo {title} {{Probing
  confinement in a $\mathbb{Z}_2$ lattice gauge theory on a quantum
  computer}},}\ }\href@noop {} {\bibfield  {journal} {\bibinfo  {journal}
  {arXiv e-prints}\ ,\ \bibinfo {eid} {arXiv:2203.08905}} (\bibinfo {year}
  {2022})},\ \Eprint {http://arxiv.org/abs/2203.08905} {arXiv:2203.08905
  [quant-ph]} \BibitemShut {NoStop}%
\bibitem [{SM()}]{SM}%
  \BibitemOpen
  \href@noop {} {}\bibinfo {howpublished} {See Supplemental Material for a
  discussion of the nonthermal steady state, supporting results for the case of
  the $\mathrm{U}(1)$ QLM with a higher link spin representation, and results
  on stabilizing and enhancing temperature-induced DFL.}\BibitemShut {Stop}%
\bibitem [{\citenamefont {Halimeh}\ \emph
  {et~al.}(2021{\natexlab{b}})\citenamefont {Halimeh}, \citenamefont {Zhao},
  \citenamefont {Hauke},\ and\ \citenamefont
  {Knolle}}]{Halimeh2021stabilizingDFL}%
  \BibitemOpen
  \bibfield  {author} {\bibinfo {author} {\bibfnamefont {Jad~C.}\ \bibnamefont
  {Halimeh}}, \bibinfo {author} {\bibfnamefont {Hongzheng}\ \bibnamefont
  {Zhao}}, \bibinfo {author} {\bibfnamefont {Philipp}\ \bibnamefont {Hauke}}, \
  and\ \bibinfo {author} {\bibfnamefont {Johannes}\ \bibnamefont {Knolle}},\
  }\bibfield  {title} {\enquote {\bibinfo {title} {Stabilizing disorder-free
  localization},}\ }\href@noop {} {\  (\bibinfo {year} {2021}{\natexlab{b}})},\
  \Eprint {http://arxiv.org/abs/2111.02427} {arXiv:2111.02427
  [cond-mat.dis-nn]} \BibitemShut {NoStop}%
\bibitem [{\citenamefont {{Homeier}}\ \emph {et~al.}(2022)\citenamefont
  {{Homeier}}, \citenamefont {{Bohrdt}}, \citenamefont {{Linsel}},
  \citenamefont {{Demler}}, \citenamefont {{Halimeh}},\ and\ \citenamefont
  {{Grusdt}}}]{Homeier2022quantum}%
  \BibitemOpen
  \bibfield  {author} {\bibinfo {author} {\bibfnamefont {Lukas}\ \bibnamefont
  {{Homeier}}}, \bibinfo {author} {\bibfnamefont {Annabelle}\ \bibnamefont
  {{Bohrdt}}}, \bibinfo {author} {\bibfnamefont {Simon}\ \bibnamefont
  {{Linsel}}}, \bibinfo {author} {\bibfnamefont {Eugene}\ \bibnamefont
  {{Demler}}}, \bibinfo {author} {\bibfnamefont {Jad~C.}\ \bibnamefont
  {{Halimeh}}}, \ and\ \bibinfo {author} {\bibfnamefont {Fabian}\ \bibnamefont
  {{Grusdt}}},\ }\bibfield  {title} {\enquote {\bibinfo {title} {{Quantum
  simulation of $\mathbb{Z}_2$ lattice gauge theories with dynamical matter
  from two-body interactions in $(2+1)$D}},}\ }\href@noop {} {\bibfield
  {journal} {\bibinfo  {journal} {arXiv e-prints}\ ,\ \bibinfo {eid}
  {arXiv:2205.08541}} (\bibinfo {year} {2022})},\ \Eprint
  {http://arxiv.org/abs/2205.08541} {arXiv:2205.08541 [cond-mat.quant-gas]}
  \BibitemShut {NoStop}%
\bibitem [{\citenamefont {{Lang}}\ \emph {et~al.}(2022)\citenamefont {{Lang}},
  \citenamefont {{Hauke}}, \citenamefont {{Knolle}}, \citenamefont {{Grusdt}},\
  and\ \citenamefont {{Halimeh}}}]{Lang2022stark}%
  \BibitemOpen
  \bibfield  {author} {\bibinfo {author} {\bibfnamefont {Haifeng}\ \bibnamefont
  {{Lang}}}, \bibinfo {author} {\bibfnamefont {Philipp}\ \bibnamefont
  {{Hauke}}}, \bibinfo {author} {\bibfnamefont {Johannes}\ \bibnamefont
  {{Knolle}}}, \bibinfo {author} {\bibfnamefont {Fabian}\ \bibnamefont
  {{Grusdt}}}, \ and\ \bibinfo {author} {\bibfnamefont {Jad~C.}\ \bibnamefont
  {{Halimeh}}},\ }\bibfield  {title} {\enquote {\bibinfo {title}
  {{Disorder-free localization with Stark gauge protection}},}\ }\href@noop {}
  {\bibfield  {journal} {\bibinfo  {journal} {arXiv e-prints}\ ,\ \bibinfo
  {eid} {arXiv:2203.01338}} (\bibinfo {year} {2022})},\ \Eprint
  {http://arxiv.org/abs/2203.01338} {arXiv:2203.01338 [cond-mat.quant-gas]}
  \BibitemShut {NoStop}%
\bibitem [{\citenamefont {Mil}\ \emph {et~al.}(2020)\citenamefont {Mil},
  \citenamefont {Zache}, \citenamefont {Hegde}, \citenamefont {Xia},
  \citenamefont {Bhatt}, \citenamefont {Oberthaler}, \citenamefont {Hauke},
  \citenamefont {Berges},\ and\ \citenamefont {Jendrzejewski}}]{Mil2020}%
  \BibitemOpen
  \bibfield  {author} {\bibinfo {author} {\bibfnamefont {Alexander}\
  \bibnamefont {Mil}}, \bibinfo {author} {\bibfnamefont {Torsten~V.}\
  \bibnamefont {Zache}}, \bibinfo {author} {\bibfnamefont {Apoorva}\
  \bibnamefont {Hegde}}, \bibinfo {author} {\bibfnamefont {Andy}\ \bibnamefont
  {Xia}}, \bibinfo {author} {\bibfnamefont {Rohit~P.}\ \bibnamefont {Bhatt}},
  \bibinfo {author} {\bibfnamefont {Markus~K.}\ \bibnamefont {Oberthaler}},
  \bibinfo {author} {\bibfnamefont {Philipp}\ \bibnamefont {Hauke}}, \bibinfo
  {author} {\bibfnamefont {J{\"u}rgen}\ \bibnamefont {Berges}}, \ and\ \bibinfo
  {author} {\bibfnamefont {Fred}\ \bibnamefont {Jendrzejewski}},\ }\bibfield
  {title} {\enquote {\bibinfo {title} {A scalable realization of local u(1)
  gauge invariance in cold atomic mixtures},}\ }\href {\doibase
  10.1126/science.aaz5312} {\bibfield  {journal} {\bibinfo  {journal}
  {Science}\ }\textbf {\bibinfo {volume} {367}},\ \bibinfo {pages} {1128--1130}
  (\bibinfo {year} {2020})}\BibitemShut {NoStop}%
\bibitem [{\citenamefont {Arute}\ \emph {et~al.}(2019)\citenamefont {Arute},
  \citenamefont {Arya}, \citenamefont {Babbush}, \citenamefont {Bacon},
  \citenamefont {Bardin}, \citenamefont {Barends}, \citenamefont {Biswas},
  \citenamefont {Boixo}, \citenamefont {Brandao}, \citenamefont {Buell},
  \citenamefont {Burkett}, \citenamefont {Chen}, \citenamefont {Chen},
  \citenamefont {Chiaro}, \citenamefont {Collins}, \citenamefont {Courtney},
  \citenamefont {Dunsworth}, \citenamefont {Farhi}, \citenamefont {Foxen},
  \citenamefont {Fowler}, \citenamefont {Gidney}, \citenamefont {Giustina},
  \citenamefont {Graff}, \citenamefont {Guerin}, \citenamefont {Habegger},
  \citenamefont {Harrigan}, \citenamefont {Hartmann}, \citenamefont {Ho},
  \citenamefont {Hoffmann}, \citenamefont {Huang}, \citenamefont {Humble},
  \citenamefont {Isakov}, \citenamefont {Jeffrey}, \citenamefont {Jiang},
  \citenamefont {Kafri}, \citenamefont {Kechedzhi}, \citenamefont {Kelly},
  \citenamefont {Klimov}, \citenamefont {Knysh}, \citenamefont {Korotkov},
  \citenamefont {Kostritsa}, \citenamefont {Landhuis}, \citenamefont
  {Lindmark}, \citenamefont {Lucero}, \citenamefont {Lyakh}, \citenamefont
  {Mandr{\`a}}, \citenamefont {McClean}, \citenamefont {McEwen}, \citenamefont
  {Megrant}, \citenamefont {Mi}, \citenamefont {Michielsen}, \citenamefont
  {Mohseni}, \citenamefont {Mutus}, \citenamefont {Naaman}, \citenamefont
  {Neeley}, \citenamefont {Neill}, \citenamefont {Niu}, \citenamefont {Ostby},
  \citenamefont {Petukhov}, \citenamefont {Platt}, \citenamefont {Quintana},
  \citenamefont {Rieffel}, \citenamefont {Roushan}, \citenamefont {Rubin},
  \citenamefont {Sank}, \citenamefont {Satzinger}, \citenamefont {Smelyanskiy},
  \citenamefont {Sung}, \citenamefont {Trevithick}, \citenamefont
  {Vainsencher}, \citenamefont {Villalonga}, \citenamefont {White},
  \citenamefont {Yao}, \citenamefont {Yeh}, \citenamefont {Zalcman},
  \citenamefont {Neven},\ and\ \citenamefont {Martinis}}]{Arute2019}%
  \BibitemOpen
  \bibfield  {author} {\bibinfo {author} {\bibfnamefont {Frank}\ \bibnamefont
  {Arute}}, \bibinfo {author} {\bibfnamefont {Kunal}\ \bibnamefont {Arya}},
  \bibinfo {author} {\bibfnamefont {Ryan}\ \bibnamefont {Babbush}}, \bibinfo
  {author} {\bibfnamefont {Dave}\ \bibnamefont {Bacon}}, \bibinfo {author}
  {\bibfnamefont {Joseph~C.}\ \bibnamefont {Bardin}}, \bibinfo {author}
  {\bibfnamefont {Rami}\ \bibnamefont {Barends}}, \bibinfo {author}
  {\bibfnamefont {Rupak}\ \bibnamefont {Biswas}}, \bibinfo {author}
  {\bibfnamefont {Sergio}\ \bibnamefont {Boixo}}, \bibinfo {author}
  {\bibfnamefont {Fernando G. S.~L.}\ \bibnamefont {Brandao}}, \bibinfo
  {author} {\bibfnamefont {David~A.}\ \bibnamefont {Buell}}, \bibinfo {author}
  {\bibfnamefont {Brian}\ \bibnamefont {Burkett}}, \bibinfo {author}
  {\bibfnamefont {Yu}~\bibnamefont {Chen}}, \bibinfo {author} {\bibfnamefont
  {Zijun}\ \bibnamefont {Chen}}, \bibinfo {author} {\bibfnamefont {Ben}\
  \bibnamefont {Chiaro}}, \bibinfo {author} {\bibfnamefont {Roberto}\
  \bibnamefont {Collins}}, \bibinfo {author} {\bibfnamefont {William}\
  \bibnamefont {Courtney}}, \bibinfo {author} {\bibfnamefont {Andrew}\
  \bibnamefont {Dunsworth}}, \bibinfo {author} {\bibfnamefont {Edward}\
  \bibnamefont {Farhi}}, \bibinfo {author} {\bibfnamefont {Brooks}\
  \bibnamefont {Foxen}}, \bibinfo {author} {\bibfnamefont {Austin}\
  \bibnamefont {Fowler}}, \bibinfo {author} {\bibfnamefont {Craig}\
  \bibnamefont {Gidney}}, \bibinfo {author} {\bibfnamefont {Marissa}\
  \bibnamefont {Giustina}}, \bibinfo {author} {\bibfnamefont {Rob}\
  \bibnamefont {Graff}}, \bibinfo {author} {\bibfnamefont {Keith}\ \bibnamefont
  {Guerin}}, \bibinfo {author} {\bibfnamefont {Steve}\ \bibnamefont
  {Habegger}}, \bibinfo {author} {\bibfnamefont {Matthew~P.}\ \bibnamefont
  {Harrigan}}, \bibinfo {author} {\bibfnamefont {Michael~J.}\ \bibnamefont
  {Hartmann}}, \bibinfo {author} {\bibfnamefont {Alan}\ \bibnamefont {Ho}},
  \bibinfo {author} {\bibfnamefont {Markus}\ \bibnamefont {Hoffmann}}, \bibinfo
  {author} {\bibfnamefont {Trent}\ \bibnamefont {Huang}}, \bibinfo {author}
  {\bibfnamefont {Travis~S.}\ \bibnamefont {Humble}}, \bibinfo {author}
  {\bibfnamefont {Sergei~V.}\ \bibnamefont {Isakov}}, \bibinfo {author}
  {\bibfnamefont {Evan}\ \bibnamefont {Jeffrey}}, \bibinfo {author}
  {\bibfnamefont {Zhang}\ \bibnamefont {Jiang}}, \bibinfo {author}
  {\bibfnamefont {Dvir}\ \bibnamefont {Kafri}}, \bibinfo {author}
  {\bibfnamefont {Kostyantyn}\ \bibnamefont {Kechedzhi}}, \bibinfo {author}
  {\bibfnamefont {Julian}\ \bibnamefont {Kelly}}, \bibinfo {author}
  {\bibfnamefont {Paul~V.}\ \bibnamefont {Klimov}}, \bibinfo {author}
  {\bibfnamefont {Sergey}\ \bibnamefont {Knysh}}, \bibinfo {author}
  {\bibfnamefont {Alexander}\ \bibnamefont {Korotkov}}, \bibinfo {author}
  {\bibfnamefont {Fedor}\ \bibnamefont {Kostritsa}}, \bibinfo {author}
  {\bibfnamefont {David}\ \bibnamefont {Landhuis}}, \bibinfo {author}
  {\bibfnamefont {Mike}\ \bibnamefont {Lindmark}}, \bibinfo {author}
  {\bibfnamefont {Erik}\ \bibnamefont {Lucero}}, \bibinfo {author}
  {\bibfnamefont {Dmitry}\ \bibnamefont {Lyakh}}, \bibinfo {author}
  {\bibfnamefont {Salvatore}\ \bibnamefont {Mandr{\`a}}}, \bibinfo {author}
  {\bibfnamefont {Jarrod~R.}\ \bibnamefont {McClean}}, \bibinfo {author}
  {\bibfnamefont {Matthew}\ \bibnamefont {McEwen}}, \bibinfo {author}
  {\bibfnamefont {Anthony}\ \bibnamefont {Megrant}}, \bibinfo {author}
  {\bibfnamefont {Xiao}\ \bibnamefont {Mi}}, \bibinfo {author} {\bibfnamefont
  {Kristel}\ \bibnamefont {Michielsen}}, \bibinfo {author} {\bibfnamefont
  {Masoud}\ \bibnamefont {Mohseni}}, \bibinfo {author} {\bibfnamefont {Josh}\
  \bibnamefont {Mutus}}, \bibinfo {author} {\bibfnamefont {Ofer}\ \bibnamefont
  {Naaman}}, \bibinfo {author} {\bibfnamefont {Matthew}\ \bibnamefont
  {Neeley}}, \bibinfo {author} {\bibfnamefont {Charles}\ \bibnamefont {Neill}},
  \bibinfo {author} {\bibfnamefont {Murphy~Yuezhen}\ \bibnamefont {Niu}},
  \bibinfo {author} {\bibfnamefont {Eric}\ \bibnamefont {Ostby}}, \bibinfo
  {author} {\bibfnamefont {Andre}\ \bibnamefont {Petukhov}}, \bibinfo {author}
  {\bibfnamefont {John~C.}\ \bibnamefont {Platt}}, \bibinfo {author}
  {\bibfnamefont {Chris}\ \bibnamefont {Quintana}}, \bibinfo {author}
  {\bibfnamefont {Eleanor~G.}\ \bibnamefont {Rieffel}}, \bibinfo {author}
  {\bibfnamefont {Pedram}\ \bibnamefont {Roushan}}, \bibinfo {author}
  {\bibfnamefont {Nicholas~C.}\ \bibnamefont {Rubin}}, \bibinfo {author}
  {\bibfnamefont {Daniel}\ \bibnamefont {Sank}}, \bibinfo {author}
  {\bibfnamefont {Kevin~J.}\ \bibnamefont {Satzinger}}, \bibinfo {author}
  {\bibfnamefont {Vadim}\ \bibnamefont {Smelyanskiy}}, \bibinfo {author}
  {\bibfnamefont {Kevin~J.}\ \bibnamefont {Sung}}, \bibinfo {author}
  {\bibfnamefont {Matthew~D.}\ \bibnamefont {Trevithick}}, \bibinfo {author}
  {\bibfnamefont {Amit}\ \bibnamefont {Vainsencher}}, \bibinfo {author}
  {\bibfnamefont {Benjamin}\ \bibnamefont {Villalonga}}, \bibinfo {author}
  {\bibfnamefont {Theodore}\ \bibnamefont {White}}, \bibinfo {author}
  {\bibfnamefont {Z.~Jamie}\ \bibnamefont {Yao}}, \bibinfo {author}
  {\bibfnamefont {Ping}\ \bibnamefont {Yeh}}, \bibinfo {author} {\bibfnamefont
  {Adam}\ \bibnamefont {Zalcman}}, \bibinfo {author} {\bibfnamefont {Hartmut}\
  \bibnamefont {Neven}}, \ and\ \bibinfo {author} {\bibfnamefont {John~M.}\
  \bibnamefont {Martinis}},\ }\bibfield  {title} {\enquote {\bibinfo {title}
  {Quantum supremacy using a programmable superconducting processor},}\ }\href
  {\doibase 10.1038/s41586-019-1666-5} {\bibfield  {journal} {\bibinfo
  {journal} {Nature}\ }\textbf {\bibinfo {volume} {574}},\ \bibinfo {pages}
  {505--510} (\bibinfo {year} {2019})}\BibitemShut {NoStop}%
\end{thebibliography}

\begin{thebibliography}{14}
\expandafter\ifx\csname natexlab\endcsname\relax\def\natexlab#1{#1}\fi
\expandafter\ifx\csname bibnamefont\endcsname\relax
  \def\bibnamefont#1{#1}\fi
\expandafter\ifx\csname bibfnamefont\endcsname\relax
  \def\bibfnamefont#1{#1}\fi
\expandafter\ifx\csname citenamefont\endcsname\relax
  \def\citenamefont#1{#1}\fi
\expandafter\ifx\csname url\endcsname\relax
  \def\url#1{\texttt{#1}}\fi
\expandafter\ifx\csname urlprefix\endcsname\relax\def\urlprefix{URL }\fi
\providecommand{\bibinfo}[2]{#2}
\providecommand{\eprint}[2][]{\url{#2}}

\bibitem[{\citenamefont{Rigol et~al.}(2008)\citenamefont{Rigol, Dunjko, and
  Olshanii}}]{Rigol_2008-S}
\bibinfo{author}{\bibfnamefont{M.}~\bibnamefont{Rigol}},
  \bibinfo{author}{\bibfnamefont{V.}~\bibnamefont{Dunjko}}, \bibnamefont{and}
  \bibinfo{author}{\bibfnamefont{M.}~\bibnamefont{Olshanii}},
  \bibinfo{journal}{Nature} \textbf{\bibinfo{volume}{452}},
  \bibinfo{pages}{854} (\bibinfo{year}{2008}),
  \urlprefix\url{https://doi.org/10.1038/nature06838}.

\bibitem[{\citenamefont{Deutsch}(1991)}]{Deutsch1991-S}
\bibinfo{author}{\bibfnamefont{J.~M.} \bibnamefont{Deutsch}},
  \bibinfo{journal}{Phys. Rev. A} \textbf{\bibinfo{volume}{43}},
  \bibinfo{pages}{2046} (\bibinfo{year}{1991}),
  \urlprefix\url{https://link.aps.org/doi/10.1103/PhysRevA.43.2046}.

\bibitem[{\citenamefont{Srednicki}(1994)}]{Srednicki1994-S}
\bibinfo{author}{\bibfnamefont{M.}~\bibnamefont{Srednicki}},
  \bibinfo{journal}{Phys. Rev. E} \textbf{\bibinfo{volume}{50}},
  \bibinfo{pages}{888} (\bibinfo{year}{1994}),
  \urlprefix\url{https://link.aps.org/doi/10.1103/PhysRevE.50.888}.

\bibitem[{\citenamefont{D'Alessio et~al.}(2016)\citenamefont{D'Alessio, Kafri,
  Polkovnikov, and Rigol}}]{Rigol_review-S}
\bibinfo{author}{\bibfnamefont{L.}~\bibnamefont{D'Alessio}},
  \bibinfo{author}{\bibfnamefont{Y.}~\bibnamefont{Kafri}},
  \bibinfo{author}{\bibfnamefont{A.}~\bibnamefont{Polkovnikov}},
  \bibnamefont{and} \bibinfo{author}{\bibfnamefont{M.}~\bibnamefont{Rigol}},
  \bibinfo{journal}{Advances in Physics} \textbf{\bibinfo{volume}{65}},
  \bibinfo{pages}{239} (\bibinfo{year}{2016}),
  \eprint{https://doi.org/10.1080/00018732.2016.1198134},
  \urlprefix\url{https://doi.org/10.1080/00018732.2016.1198134}.

\bibitem[{\citenamefont{Smith et~al.}(2018)\citenamefont{Smith, Knolle,
  Moessner, and Kovrizhin}}]{Smith2018-S}
\bibinfo{author}{\bibfnamefont{A.}~\bibnamefont{Smith}},
  \bibinfo{author}{\bibfnamefont{J.}~\bibnamefont{Knolle}},
  \bibinfo{author}{\bibfnamefont{R.}~\bibnamefont{Moessner}}, \bibnamefont{and}
  \bibinfo{author}{\bibfnamefont{D.~L.} \bibnamefont{Kovrizhin}},
  \bibinfo{journal}{Phys. Rev. B} \textbf{\bibinfo{volume}{97}},
  \bibinfo{pages}{245137} (\bibinfo{year}{2018}),
  \urlprefix\url{https://link.aps.org/doi/10.1103/PhysRevB.97.245137}.

\bibitem[{\citenamefont{Halimeh
  et~al.}(2021{\natexlab{a}})\citenamefont{Halimeh, Lang, Mildenberger, Jiang,
  and Hauke}}]{Halimeh2020e-S}
\bibinfo{author}{\bibfnamefont{J.~C.} \bibnamefont{Halimeh}},
  \bibinfo{author}{\bibfnamefont{H.}~\bibnamefont{Lang}},
  \bibinfo{author}{\bibfnamefont{J.}~\bibnamefont{Mildenberger}},
  \bibinfo{author}{\bibfnamefont{Z.}~\bibnamefont{Jiang}}, \bibnamefont{and}
  \bibinfo{author}{\bibfnamefont{P.}~\bibnamefont{Hauke}},
  \bibinfo{journal}{PRX Quantum} \textbf{\bibinfo{volume}{2}},
  \bibinfo{pages}{040311} (\bibinfo{year}{2021}{\natexlab{a}}),
  \urlprefix\url{https://link.aps.org/doi/10.1103/PRXQuantum.2.040311}.

\bibitem[{\citenamefont{Halimeh
  et~al.}(2021{\natexlab{b}})\citenamefont{Halimeh, Homeier, Schweizer,
  Aidelsburger, Hauke, and Grusdt}}]{Halimeh2021stabilizing-S}
\bibinfo{author}{\bibfnamefont{J.~C.} \bibnamefont{Halimeh}},
  \bibinfo{author}{\bibfnamefont{L.}~\bibnamefont{Homeier}},
  \bibinfo{author}{\bibfnamefont{C.}~\bibnamefont{Schweizer}},
  \bibinfo{author}{\bibfnamefont{M.}~\bibnamefont{Aidelsburger}},
  \bibinfo{author}{\bibfnamefont{P.}~\bibnamefont{Hauke}}, \bibnamefont{and}
  \bibinfo{author}{\bibfnamefont{F.}~\bibnamefont{Grusdt}}
  (\bibinfo{year}{2021}{\natexlab{b}}), \eprint{2108.02203}.

\bibitem[{\citenamefont{Halimeh
  et~al.}(2021{\natexlab{c}})\citenamefont{Halimeh, Zhao, Hauke, and
  Knolle}}]{Halimeh2021stabilizingDFL-S}
\bibinfo{author}{\bibfnamefont{J.~C.} \bibnamefont{Halimeh}},
  \bibinfo{author}{\bibfnamefont{H.}~\bibnamefont{Zhao}},
  \bibinfo{author}{\bibfnamefont{P.}~\bibnamefont{Hauke}}, \bibnamefont{and}
  \bibinfo{author}{\bibfnamefont{J.}~\bibnamefont{Knolle}}
  (\bibinfo{year}{2021}{\natexlab{c}}), \eprint{2111.02427}.

\bibitem[{\citenamefont{Halimeh et~al.}(2022)\citenamefont{Halimeh, Homeier,
  Zhao, Bohrdt, Grusdt, Hauke, and Knolle}}]{Halimeh2021enhancing-S}
\bibinfo{author}{\bibfnamefont{J.~C.} \bibnamefont{Halimeh}},
  \bibinfo{author}{\bibfnamefont{L.}~\bibnamefont{Homeier}},
  \bibinfo{author}{\bibfnamefont{H.}~\bibnamefont{Zhao}},
  \bibinfo{author}{\bibfnamefont{A.}~\bibnamefont{Bohrdt}},
  \bibinfo{author}{\bibfnamefont{F.}~\bibnamefont{Grusdt}},
  \bibinfo{author}{\bibfnamefont{P.}~\bibnamefont{Hauke}}, \bibnamefont{and}
  \bibinfo{author}{\bibfnamefont{J.}~\bibnamefont{Knolle}},
  \bibinfo{journal}{PRX Quantum} \textbf{\bibinfo{volume}{3}},
  \bibinfo{pages}{020345} (\bibinfo{year}{2022}),
  \urlprefix\url{https://link.aps.org/doi/10.1103/PRXQuantum.3.020345}.

\bibitem[{\citenamefont{{Lang} et~al.}(2022)\citenamefont{{Lang}, {Hauke},
  {Knolle}, {Grusdt}, and {Halimeh}}}]{Lang2022stark-S}
\bibinfo{author}{\bibfnamefont{H.}~\bibnamefont{{Lang}}},
  \bibinfo{author}{\bibfnamefont{P.}~\bibnamefont{{Hauke}}},
  \bibinfo{author}{\bibfnamefont{J.}~\bibnamefont{{Knolle}}},
  \bibinfo{author}{\bibfnamefont{F.}~\bibnamefont{{Grusdt}}}, \bibnamefont{and}
  \bibinfo{author}{\bibfnamefont{J.~C.} \bibnamefont{{Halimeh}}},
  \bibinfo{journal}{arXiv e-prints} \bibinfo{eid}{arXiv:2203.01338}
  (\bibinfo{year}{2022}), \eprint{2203.01338}.

\bibitem[{\citenamefont{Schweizer et~al.}(2019)\citenamefont{Schweizer, Grusdt,
  Berngruber, Barbiero, Demler, Goldman, Bloch, and
  Aidelsburger}}]{Schweizer2019-S}
\bibinfo{author}{\bibfnamefont{C.}~\bibnamefont{Schweizer}},
  \bibinfo{author}{\bibfnamefont{F.}~\bibnamefont{Grusdt}},
  \bibinfo{author}{\bibfnamefont{M.}~\bibnamefont{Berngruber}},
  \bibinfo{author}{\bibfnamefont{L.}~\bibnamefont{Barbiero}},
  \bibinfo{author}{\bibfnamefont{E.}~\bibnamefont{Demler}},
  \bibinfo{author}{\bibfnamefont{N.}~\bibnamefont{Goldman}},
  \bibinfo{author}{\bibfnamefont{I.}~\bibnamefont{Bloch}}, \bibnamefont{and}
  \bibinfo{author}{\bibfnamefont{M.}~\bibnamefont{Aidelsburger}},
  \bibinfo{journal}{Nature Physics} \textbf{\bibinfo{volume}{15}},
  \bibinfo{pages}{1168} (\bibinfo{year}{2019}),
  \urlprefix\url{https://doi.org/10.1038/s41567-019-0649-7}.

\bibitem[{\citenamefont{Mil et~al.}(2020)\citenamefont{Mil, Zache, Hegde, Xia,
  Bhatt, Oberthaler, Hauke, Berges, and Jendrzejewski}}]{Mil2020-S}
\bibinfo{author}{\bibfnamefont{A.}~\bibnamefont{Mil}},
  \bibinfo{author}{\bibfnamefont{T.~V.} \bibnamefont{Zache}},
  \bibinfo{author}{\bibfnamefont{A.}~\bibnamefont{Hegde}},
  \bibinfo{author}{\bibfnamefont{A.}~\bibnamefont{Xia}},
  \bibinfo{author}{\bibfnamefont{R.~P.} \bibnamefont{Bhatt}},
  \bibinfo{author}{\bibfnamefont{M.~K.} \bibnamefont{Oberthaler}},
  \bibinfo{author}{\bibfnamefont{P.}~\bibnamefont{Hauke}},
  \bibinfo{author}{\bibfnamefont{J.}~\bibnamefont{Berges}}, \bibnamefont{and}
  \bibinfo{author}{\bibfnamefont{F.}~\bibnamefont{Jendrzejewski}},
  \bibinfo{journal}{Science} \textbf{\bibinfo{volume}{367}},
  \bibinfo{pages}{1128} (\bibinfo{year}{2020}), ISSN \bibinfo{issn}{0036-8075},
  \urlprefix\url{https://science.sciencemag.org/content/367/6482/1128}.

\bibitem[{\citenamefont{Yang et~al.}(2020)\citenamefont{Yang, Sun, Ott, Wang,
  Zache, Halimeh, Yuan, Hauke, and Pan}}]{Yang2020-S}
\bibinfo{author}{\bibfnamefont{B.}~\bibnamefont{Yang}},
  \bibinfo{author}{\bibfnamefont{H.}~\bibnamefont{Sun}},
  \bibinfo{author}{\bibfnamefont{R.}~\bibnamefont{Ott}},
  \bibinfo{author}{\bibfnamefont{H.-Y.} \bibnamefont{Wang}},
  \bibinfo{author}{\bibfnamefont{T.~V.} \bibnamefont{Zache}},
  \bibinfo{author}{\bibfnamefont{J.~C.} \bibnamefont{Halimeh}},
  \bibinfo{author}{\bibfnamefont{Z.-S.} \bibnamefont{Yuan}},
  \bibinfo{author}{\bibfnamefont{P.}~\bibnamefont{Hauke}}, \bibnamefont{and}
  \bibinfo{author}{\bibfnamefont{J.-W.} \bibnamefont{Pan}},
  \bibinfo{journal}{Nature} \textbf{\bibinfo{volume}{587}},
  \bibinfo{pages}{392} (\bibinfo{year}{2020}),
  \urlprefix\url{https://doi.org/10.1038/s41586-020-2910-8}.

\bibitem[{\citenamefont{Zhou et~al.}(2021)\citenamefont{Zhou, Su, Halimeh, Ott,
  Sun, Hauke, Yang, Yuan, Berges, and Pan}}]{Zhou2021-S}
\bibinfo{author}{\bibfnamefont{Z.-Y.} \bibnamefont{Zhou}},
  \bibinfo{author}{\bibfnamefont{G.-X.} \bibnamefont{Su}},
  \bibinfo{author}{\bibfnamefont{J.~C.} \bibnamefont{Halimeh}},
  \bibinfo{author}{\bibfnamefont{R.}~\bibnamefont{Ott}},
  \bibinfo{author}{\bibfnamefont{H.}~\bibnamefont{Sun}},
  \bibinfo{author}{\bibfnamefont{P.}~\bibnamefont{Hauke}},
  \bibinfo{author}{\bibfnamefont{B.}~\bibnamefont{Yang}},
  \bibinfo{author}{\bibfnamefont{Z.-S.} \bibnamefont{Yuan}},
  \bibinfo{author}{\bibfnamefont{J.}~\bibnamefont{Berges}}, \bibnamefont{and}
  \bibinfo{author}{\bibfnamefont{J.-W.} \bibnamefont{Pan}}
  (\bibinfo{year}{2021}), \eprint{2107.13563}.

\end{thebibliography}
\end{document}